\documentclass[letterpaper,12pt]{article}
\pdfoutput=1

\usepackage{enumerate}
\usepackage{jheppub}
\usepackage{amsmath}
\usepackage{graphicx}
\usepackage{subfig}
\usepackage{cancel}
\usepackage[dvipsnames]{xcolor}
\usepackage{color}
\usepackage{mathrsfs}
\usepackage{amssymb}
\usepackage{epstopdf}
\usepackage{dsfont}
\usepackage{float}
\def\({\left(}
\def\){\right)}
\def\[{\left[}
\def\]{\right]}

\newpage

\title{CFT sewing as the dual of AdS cut-and-paste}

\author{Donald Marolf}

\affiliation{Department of Physics, University of California, Santa Barbara, CA 93106, USA}

\emailAdd{marolf@physics.ucsb.edu}

\abstract{The CPT map allows two states of a quantum field theory to be sewn together over CPT-conjugate partial Cauchy surfaces $R_1,R_2$ to make a state on a new spacetime.  We study the holographic dual of this operation in the case where the original states are CPT-conjugate within $R_1,R_2$ to leading order in the bulk Newton constant $G$, and where the bulk duals are dominated by classical bulk geometries $g_1,g_2$.  For states of fixed area on the $R_1,R_2$ HRT-surfaces, we argue that the bulk geometry $g_1 \#  g_2$ dual to the newly sewn state is given by deleting the entanglement wedges of $R_1,R_2$ from $g_1,g_2$, gluing the remaining complementary entanglement wedges of ${\bar R}_1, {\bar R}_2$  together across the HRT surface, and solving the equations of motion to the past and future.  The argument uses the bulk path integral and assumes it to be dominated by a certain natural saddle.  For states where the HRT area is not fixed, the same bulk cut-and-paste is dual to a modified sewing that produces a generalization of the canonical purification state $\sqrt{\rho}$ discussed recently by Dutta and Faulkner.  Either form of the construction can be used to build CFT states dual to bulk geometries associated with multipartite reflected entropy.}

\begin{document}
\maketitle

\section{Introduction}
\label{Introduction}

The tensor network models of holography introduced in \cite{Swingle:2009bg} (see also e.g. \cite{Qi:2013caa,Evenbly,MolinaVilaplana:2011xt,Swingle:2012wq,Matsueda:2012xm}) motivate the idea that general codimension-2 surfaces in the bulk carry a notion of bulk quantum state.    Furthermore, in simple tensor networks that yield quantum error correcting (QEC) codes \cite{Pastawski:2015qua,Hayden:2016cfa} the network in an entanglement wedge defines an isometry between a bulk state on the Ryu-Takayanagi (RT) surface and the state of the dual CFT on the corresponding boundary region.  This provides an especially concrete model of the holographic entanglement wedge dualities proposed in \cite{Czech:2012bh,Bousso:2012mh,Hubeny:2012wa} associated with the corresponding RT and Hubeny-Rangamani-Takayanagi (HRT) conjectures \cite{Ryu:2006ef,Ryu:2006bv,Hubeny:2007xt}. Such observations have inspired interesting suggestions for extending the holographic dictionary, including the recently conjectured surface-state correspondence \cite{Miyaji:2015yva} and a variety of proposed duals \cite{Takayanagi:2017knl,Nguyen:2017yqw,Tamaoka:2018ned,Kudler-Flam:2018qjo,Dutta:2019gen} for the so-called entanglement wedge cross section; see also related studies in \cite{Umemoto:2018jpc,Bao:2017nhh,Hirai:2018jwy,Bao:2018gck,Bao:2018fso,Agon:2018lwq,Caputa:2018xuf,Kudler-Flam:2019oru,Du:2019emy,Jokela:2019ebz,Bao:2019wcf,Harper:2019lff,Kusuki:2019zsp,Levin:2019krg,Kusuki:2019evw}.

In this work, we consider implications of the QEC tensor network paradigm for the natural operation of sewing together two CFT states over CPT-conjugate regions $R_1,R_2$.
In a fully covariant discussion it is natural to use $R_1,R_2$ to denote domains of dependence in the Lorentz-signature spacetimes $M_1,M_2$ on which the CFT states are defined, and to then take $\bar R_1, \bar R_2$ to be the complementary domains of dependence such that $R_1 \cup \bar R_1$ and $R_2 \cup \bar R_2$ respectively contain Cauchy surfaces for $M_1$ and $M_2$.  But it also suffices to fix Cauchy surfaces in $M_1,M_2$ and to take $(R_1,\bar R_1)$ and $(R_2,\bar R_2)$ as pairs of complementary regions in the chosen surfaces.  To avoid lengthy definitions we will use the two descriptions interchangeably.  The latter description is particularly useful to define the boundaries $\partial R_1, \partial R_2$ of our regions, which we take to be codimension-2 surfaces in $M_1,M_2$.

\begin{figure}[h]
\centering
\includegraphics[width =0.35\textwidth]{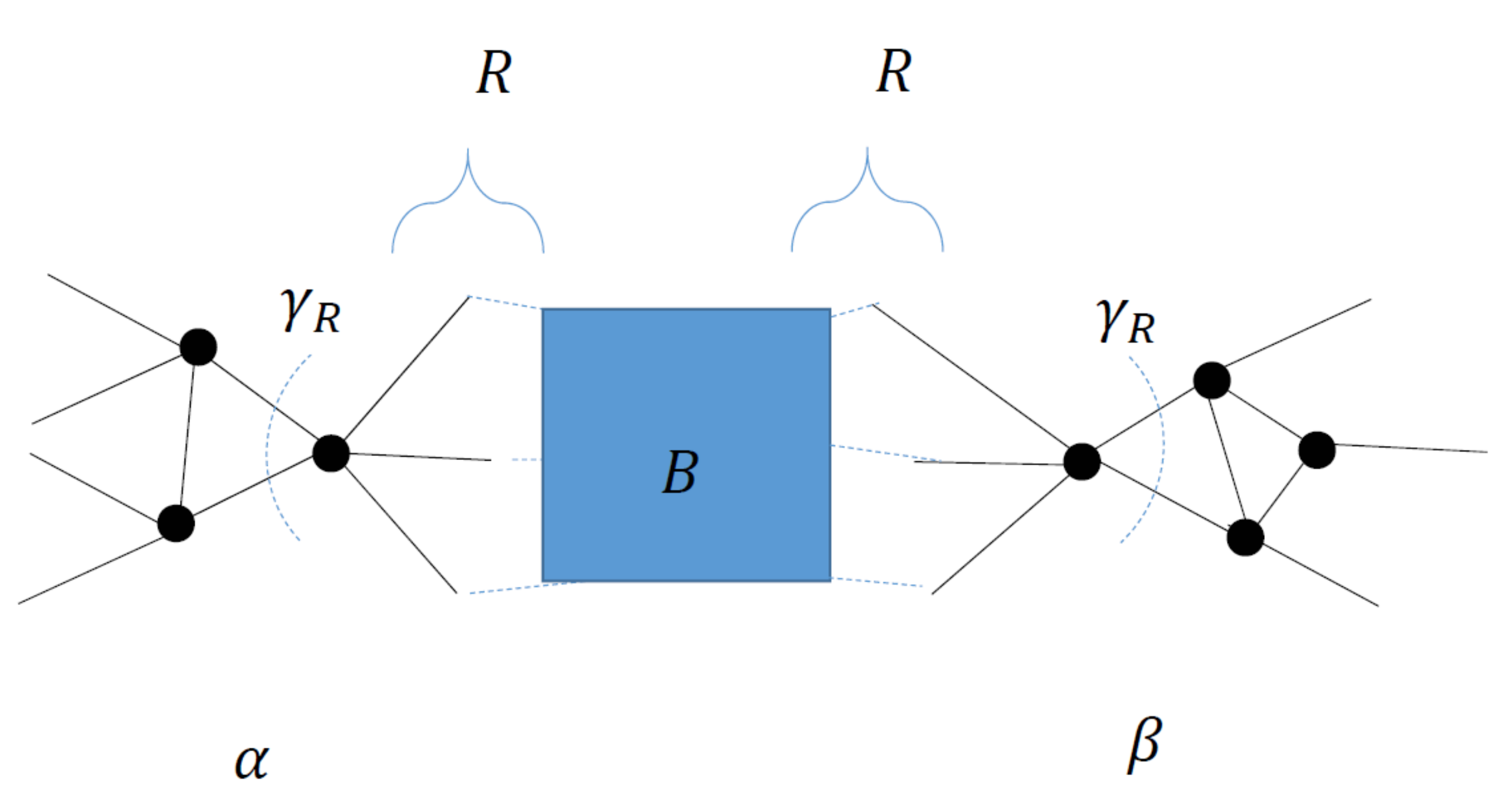}\includegraphics[width =0.35\textwidth]{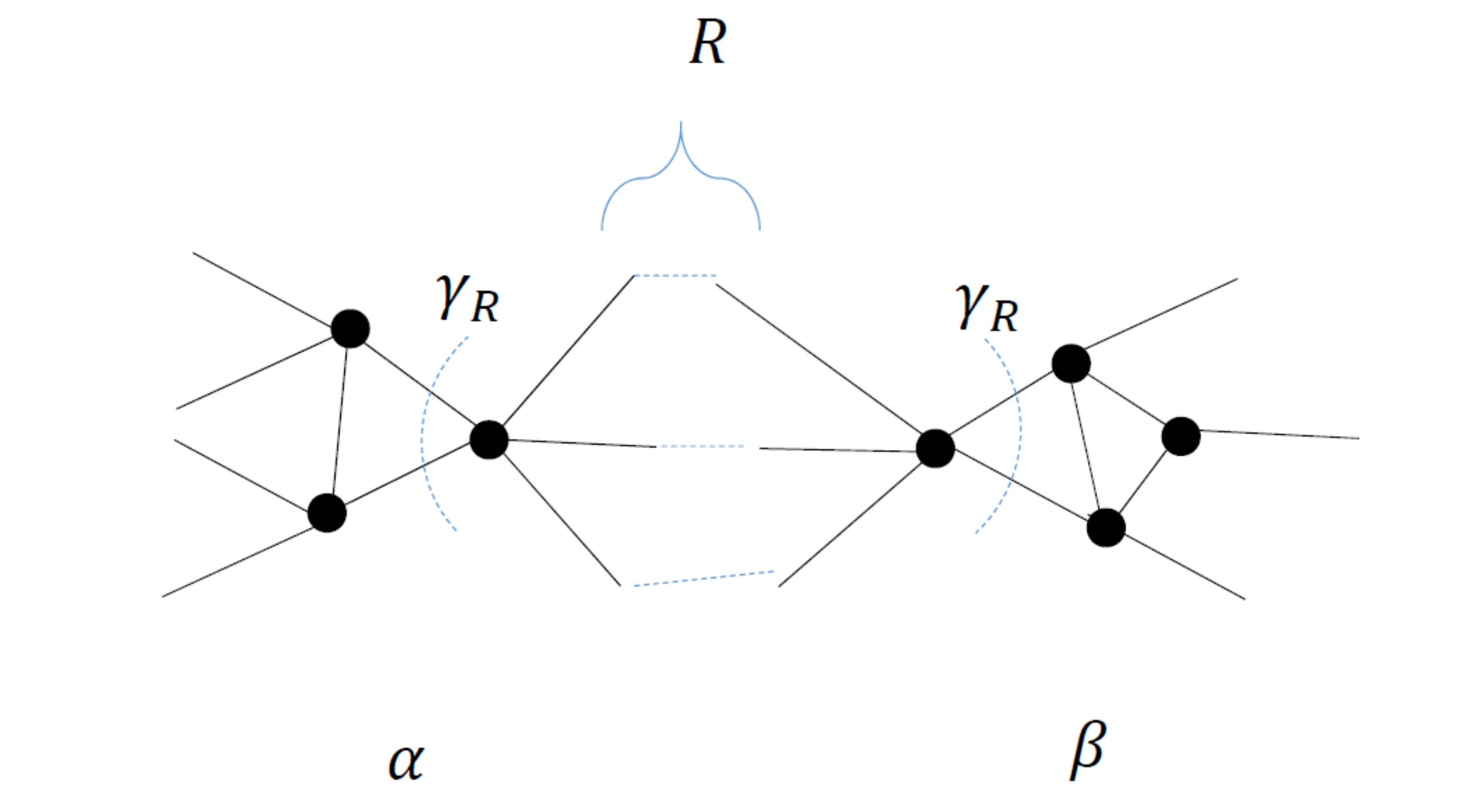}\\ \medskip \medskip
\includegraphics[width =0.35\textwidth]{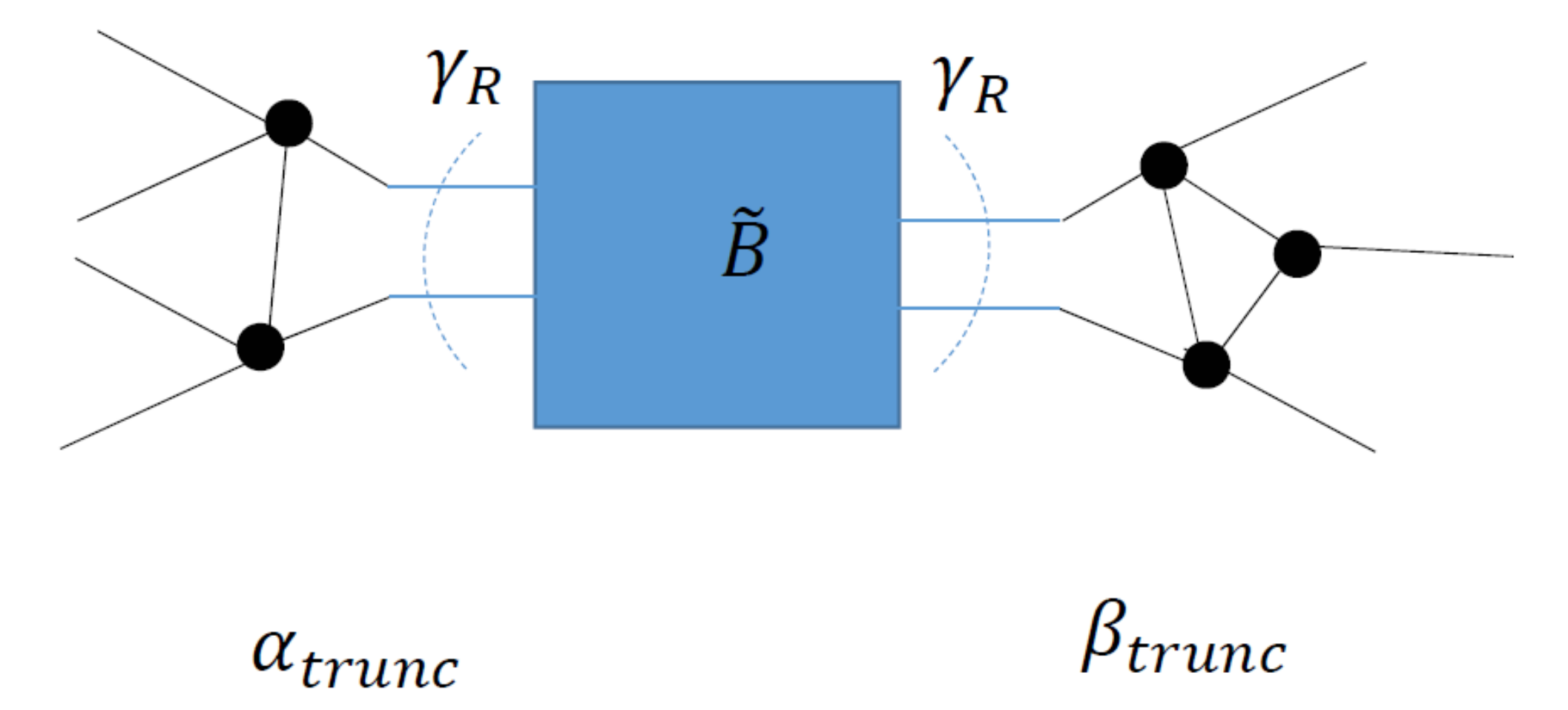}\includegraphics[width =0.35\textwidth]{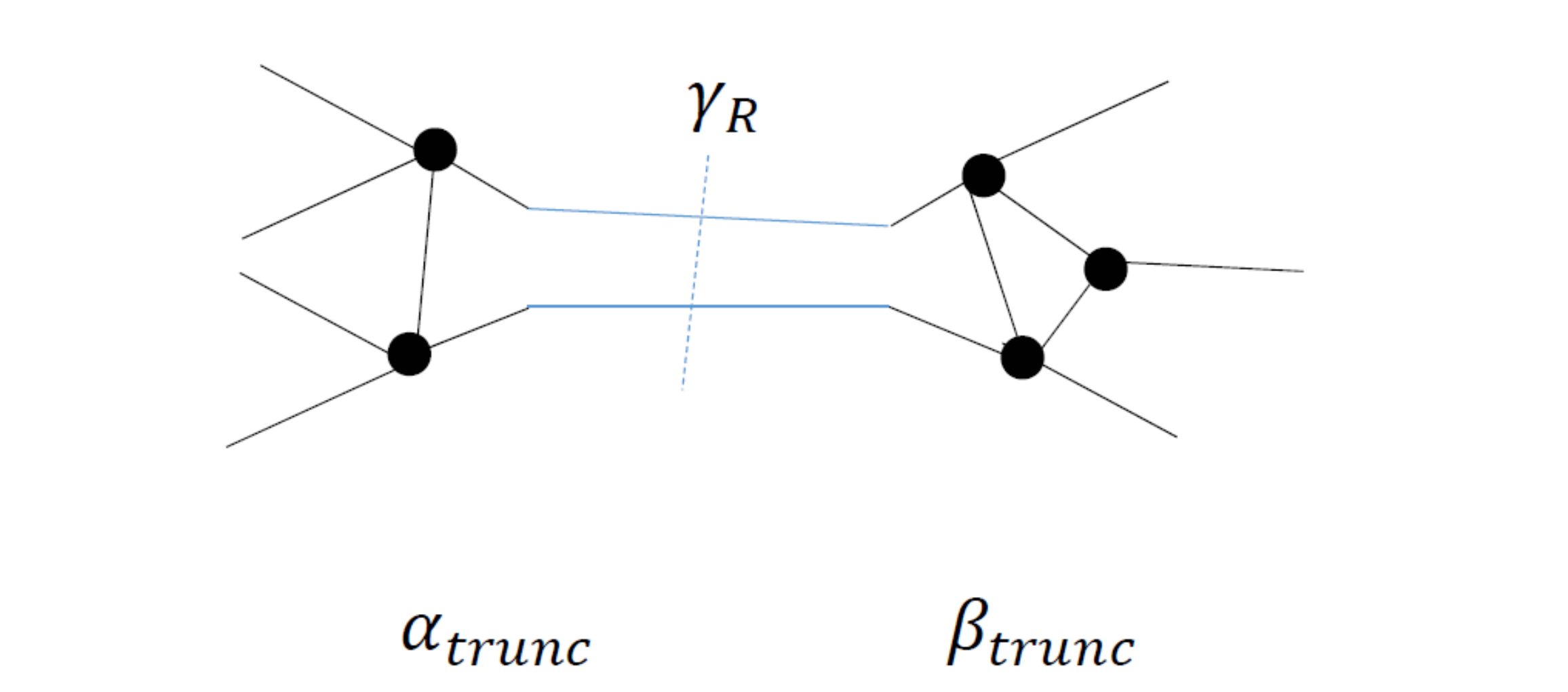}\\ \medskip \medskip
\includegraphics[width =0.32\textwidth]{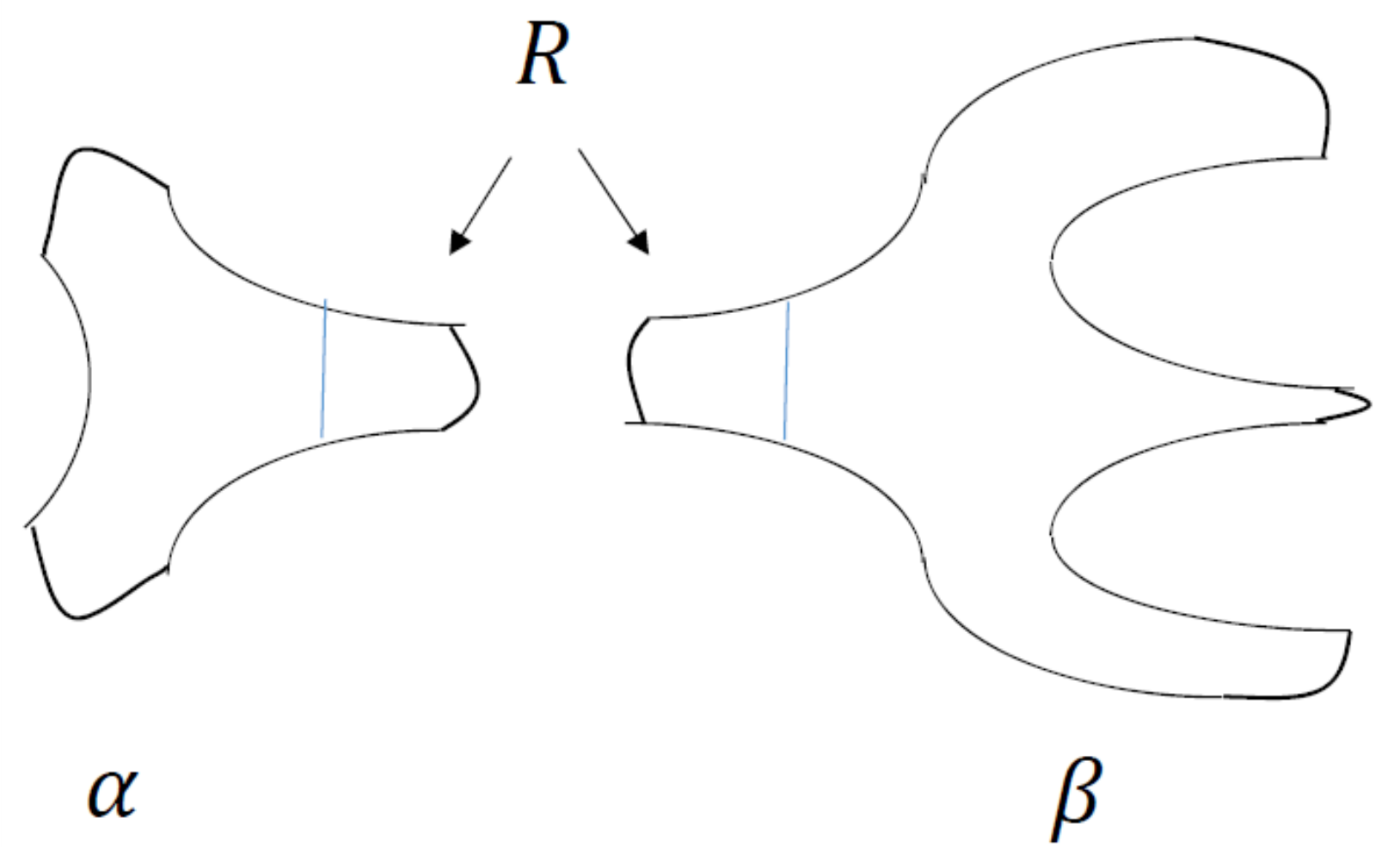} \includegraphics[width =0.35
\textwidth]{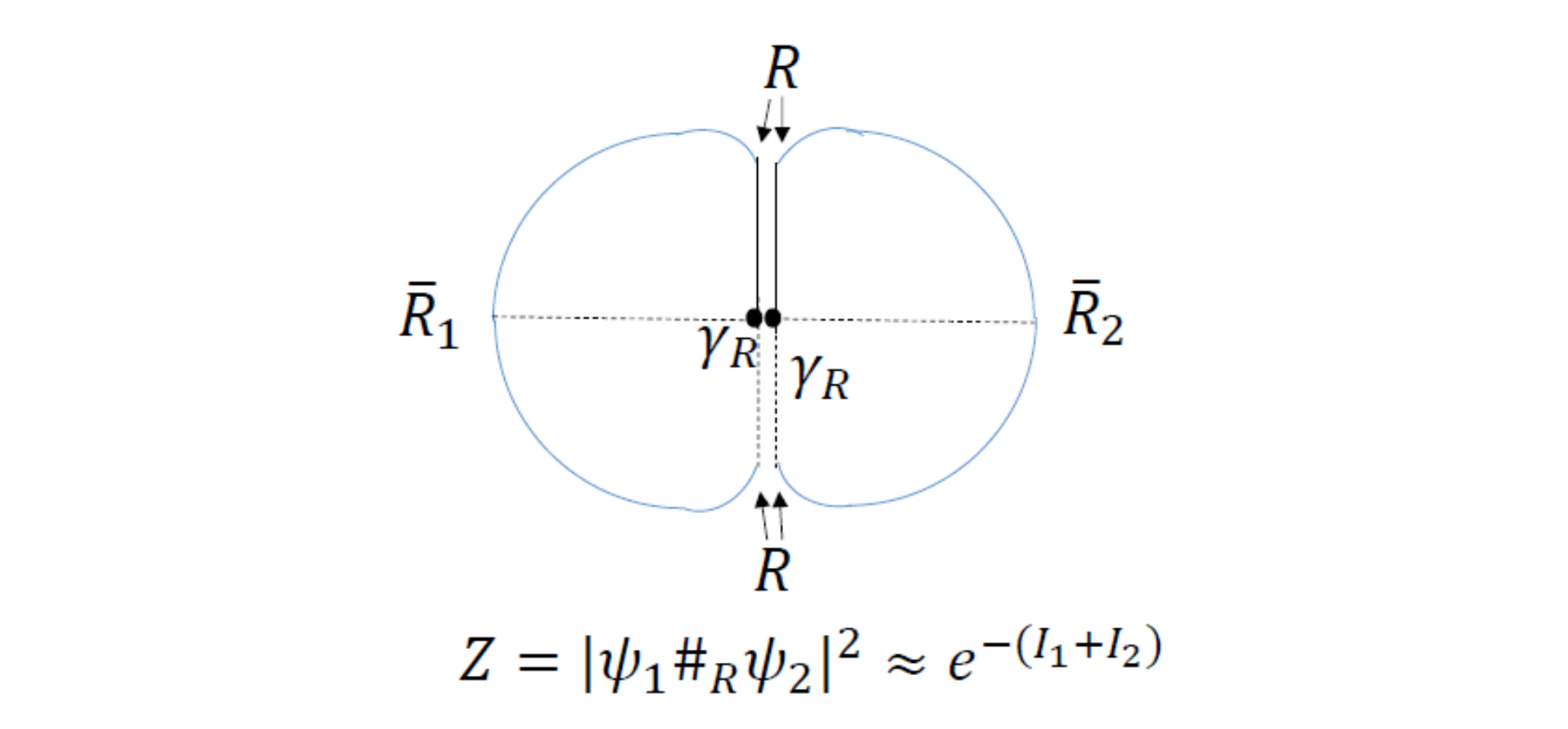}\\
\caption{Various representations of $\alpha \#_R \beta$.  {\bf Top Row:} Tensor network states $\alpha, \beta$ both have boundaries containing regions called $R$ and associated RT surfaces $\gamma_R$ (dashed arcs).  At left, the outputs in $R$ are contracted using the bilinear form $B$.  This defines  $\alpha \#_R \beta$.  If $B$ is constructed from the adjoint map and a local anti-linear CPT operation, the contraction over $R$ becomes local as shown at right.  {\bf Center Row:} The isometries $V_1,V_2$ defined by the entanglement wedges have been used to rewrite the contraction defining $\alpha \#_R \beta$ as acting on truncated tensor networks from which the entanglement wedge of $R$ has been excised.  The truncated networks define new states $\alpha_{trunc}, \beta_{trunc}$.  At left, the contraction is implemented by the bilinear form $\tilde B = B \circ (V_1 \otimes V_2)$.  If $V_1, V_2$ are CPT conjugate under a local CPT operation, this contraction becomes local on $\gamma_R$ as at right. {\bf Bottom Row:}  The analogous cut-and-paste construction for time-symmetric slices of bulk spacetimes with CPT-conjugate regions $R$. Slices of the original spacetimes are shown at left, while the result of the cut-and-paste is shown at right.}
\label{fig:contractTNs}
\end{figure}

As currently understood, tensor networks model only time-symmetric holographic states.  We thus focus on this case in our introductory motivations, taking $R_1,R_2$ be isometric in a time-orientation-preserving sense and writing $R_1 = R = R_2$.  However, we still require any sources in $R_1$ to be CPT-conjugate to those in $R_2$.

For simplicity, we assume $R$ to be associated with a factor ${\cal H}_R$ of the CFT Hilbert space ${\cal H}_{CFT} = {\cal H}_R \otimes {\cal H}_{\bar R}$; i.e., we ignore complications associated with the type III von Neumann algebras of continuum quantum field theory. We also suppose that we have chosen a bilinear form $B: \ {\cal H}_R \otimes {\cal H}_R \rightarrow {\mathbb C}$.  Equivalently, we may construct $B$ from the inner-product and a choice of anti-unitary operator that we choose to call $(CPT)_R: {\cal H}_R \rightarrow {\cal H}_R$.  

Given two states $|\alpha \rangle$, $|\beta\rangle$ defined by tensor networks $TN_1,TN_2$, we may use $B$ to contract the indices in $R$ and thus to build a new state $|\alpha \#_R \beta \rangle$ on a new boundary defined by joining together the complementary regions ${\bar R}_1, {\bar R}_2$.  Assuming that $(CPT)_R$ acts locally on each $R$-output of the network, this contraction is local as shown in the top line of figure \ref{fig:contractTNs} above.  Such locality is expected in quantum field theories, since $CPT$ must map the algebra of local operators in any spacetime region to the algebra in the $CPT$-conjugate region.  Note that we are free to consider the case where ${\bar R}_1, {\bar R}_2$ are distinct, so long as the sewing takes place over a common region $R$ of the original boundaries.  In the CFT context, this corresponds to the case where the original states $|\alpha \rangle$, $|\beta\rangle$ live on different spacetimes

We now suppose that, as in the QEC context above, each network $TN_i$ defines an isometry $V_i$ from  a Hilbert space ${\cal H}_{\gamma_{R}}$ on its RT surface $\gamma_R$ to ${\cal H}_R$.
This allows the sewn state $|\alpha \#_R \beta \rangle$ to be written in terms of two amputated tensor networks defined by removing the entanglement wedge of $R$ from each $TN_i$.  In particular, $|\alpha \#_R \beta \rangle$ is obtained by sewing the amputated networks together along the RT surface (see 2nd line in figure \ref{fig:contractTNs}) using the bilinear form
\begin{equation}
\tilde B:  \ {\cal H}_{\gamma_{R}} \otimes {\cal H}_{\gamma_{R}} \rightarrow {\mathbb C},  \ \ \ \tilde B = B \circ (V_1 \otimes V_2); \\
\end{equation}
i.e., with
\begin{equation}
\tilde B(\tilde \psi_{1}, \tilde \psi_2)  = B(V_1 \tilde\psi_1, V_2\tilde\psi_2).
\end{equation}

Furthermore, if we also have a notion of CPT on the RT surface $\gamma_{R}$ corresponding to some anti-unitary operator
$(CPT)_{\gamma_{R}}: {\cal H}_{\gamma_{R}} \rightarrow {\cal H}_{\gamma_{R}}$, our RT sewing operator $\tilde B$ takes a particularly simple form when $V_2$ is CPT-conjugate to $V_1$; i.e., when
$V_1 = (CPT)_R  V_2 (CPT)^\dagger_{\gamma_{R}}$.  To see this, let us denote the inner product on either ${\cal H}_R$ or ${\cal H}_{\gamma_{R}}$ by $ \Bigl(\phi_2, \phi_1\Bigr) = \langle \phi_2 |\phi_1 \rangle$. Then choosing
\begin{equation}
\label{eq:natural}
B(\psi_1, \psi_2) = \langle (CPT_R)\psi_2|\psi_1\rangle := \Bigl( (CPT_R)\psi_2, \psi_1 \Bigr),
\end{equation}
we find
\begin{equation}
\tilde B(\tilde\psi_1, \tilde\psi_2) = \Bigl( (CPT_R)V_2 \tilde\psi_2, V_1 \tilde\psi_1 \Bigr) =  \Bigl( (CPT_{\gamma_{R}}) \tilde\psi_2,\tilde\psi_1 \Bigr).
\end{equation}

This is just the natural contraction of indices on $\gamma_{R}$ defined by $CPT_{\gamma_{R}}$.  And just as on $R$, if $CPT_{\gamma_{R}}$ acts locally on $\gamma_{R}$ the contraction is local as shown in the second line of figure \ref{fig:contractTNs}.  The analogy with AdS/CFT would then suggest that in the time-symmetric case the CPT-sewing of two states over a region $R$ is dual to the bulk cut-and-paste procedure shown in the final line of figure \ref{fig:contractTNs}, where the initial data for the final spacetime is constructed by first removing the entanglement wedge of $R$ from both original spacetimes and then sewing the remainders together along the RT surface, or more generally along the HRT surface in cases without time-symmetry.

Note that this initial data will generally not be smooth.  While taking the limit at $\gamma_R$ from the $R$ entanglement wedges shows that the fields are continuous at $\gamma_R$, their normal derivatives typically change signs across this surface.  As a result, as in the holographic canonical purification of \cite{Dutta:2019gen}, applying our cut-and-paste to smooth bulk solutions generally leads to a spacetime with a mild shockwave (representing a discontinuity, but not a divergence, in the shear or matter stress tensor) propagating along both null surfaces orthogonal ato $\gamma_{R}$.

In considering this analogy, it is important to recall that
while the tensor networks of \cite{Pastawski:2015qua,Hayden:2016cfa} model certain aspects of holography, the properties of their Renyi entropies $S_n$ differ markedly from those of general holographic states (see e.g. \cite{Hayden:2016cfa}). Renyi entropies involve powers of a density matrix $\rho$, and computing e.g. $\rho^2$ has much in common with the above sewing of CFT states.  So one would expect similarly general failures of the above analogy. However, as described in \cite{Akers:2018fow,Dong:2018seb}, for a given choice of region $R$ such tensor networks {\it do} correctly model the Renyi entropies of holographic states defined by first fixing the area of the HRT surface $\gamma_R = \gamma_{\bar R}$. More general states can then be described by superpositions, and thus can instead be modeled using the edge-mode tensor networks described in \cite{Donnelly:2016qqt}, but for simplicity we first focus on fixed-area states below.

Continuing to use language appropriate to the time-symmetric case, 
we thus expect holographic systems to display a duality between sewing fixed-area versions of CFT states along $R$ and the construction of a new bulk spacetime by gluing together the entanglement wedges of $\bar R_1$ and $\bar R_2$; see again the final line of figure \ref{fig:contractTNs}.   We use the bulk path integral to give a simple argument for this duality below. In parallel with related assumptions in \cite{Dong:2018seb,Dutta:2019gen} and to some extent also in \cite{Lewkowycz:2013nqa}, this argument assumes that a natural saddle dominates the path integral.  For simplicity we work with Euclidean path integrals and time-symmetric states, but as usual the arguments extend to more general time-dependent cases by upgrading Euclidean path integrals to Schwinger-Keldysh path integrals as in \cite{Dong:2016hjy}.

In the special case where the original fixed-area states $|\alpha\rangle, |\beta\rangle$ are CPT-conjugate on the entire spacetime, the final state is just the canonical purifcation described in \cite{Dutta:2019gen} of the density matrix induced on ${\bar R}_1$ (or equivalently on ${\bar R}_2$). Indeed, after understanding the fixed-area case it will be clear at leading order in $G$ that a similar duality holds for arbitrary CFT states, so long as their bulk duals are dominated by a single classical geometry.  This more general duality uses the same bulk cut-and-paste, but involves a modified CFT sewing procedure. Our construction thus generalizes that of \cite{Dutta:2019gen}, and in particular can be used to construct CFT states dual to the cut-and-paste geometries of  \cite{Bao:2019wcf} associated with reflected multipartite entanglement.

We begin in section \ref{review} with a brief review of fixed area states.  The main argument follows in section \ref{arg}. We close with some brief discussion in section \ref{disc}, describing the modified CFT sewing needed for states whose HRT-areas have not been fixed as well as another generalization that involves sewing a given state to iteself.

\section{Fixed-Area State Review}
\label{review}

We are interested in CFT states prepared by path integrals and their dual bulk representations.  For a given CFT region $R$ and its complementary $\bar R$ region, there is a diffeomorphsim-invariant bulk operator $A(\gamma_{\bar R})$ that gives the area of the associated HRT surface $\gamma_{\bar R}=\gamma_R$.  At least at the level of semi-classical bulk physics, this $A(\gamma_{\bar R})$ is self-adjoint and we can imagine decomposing any state $|\psi\rangle$ into eigenstates of $A(\gamma_{\bar R})$, or into approximate eigenstates smeared over small regions of the spectrum of $A(\gamma_{\bar R})$:
\begin{equation}
|\psi \rangle = \sum_A |\psi; A \rangle.
\end{equation}
As described in \cite{Akers:2018fow,Dong:2018seb} this decomposition is deeply related to the quantum error correction (QEC) structure \cite{Harlow:2016vwg} of CFT states with respect to $R$ and the complementary region $\bar R$.  In particular, $A(\gamma_{\bar R})$ is a central element of the algebra recovered by quantum error correction.

However, we will not need the details of the relation to QEC here.  Indeed, all we really need for the argument in section \ref{arg} are the following results (see e.g. \cite{Dong:2018seb}):

\begin{enumerate}[(i)]

\item{}$A(\gamma_{\bar R})$ can be reconstructed in either CFT region $R$ or $\bar R$.

\item{} At least at leading order in the bulk Newton constant $G$, the probability finding some approximate value for $A$ is given by evaluating the path integral for the norm $\langle \psi | \psi \rangle$ with an extra constraint fixing the value of $A = A(\gamma_{\bar R})$.  Here as in \cite{Dong:2018seb} seek only to fix $A$ up to $O(G)$ corrections and not to define an exact area-eigenstate.

\item{} Since the constraint in (ii) prohibits us from integrating over one function of the bulk metric, saddle points of the constrained path integral need not satisfy the corresponding equation of motion. In Euclidean signature, it turns out that saddle points for the path integral with fixed $A(\gamma_{\bar R})$ can include a codimension-2 conical defect anchored at the boundary to $\partial R$.  As in \cite{Lewkowycz:2013nqa}, one may think of this as a spacelike codimension-2 cosmic-brane anchored to the boundary at $\partial R$.  This brane describes $\gamma_{\bar R}$ for the fixed-area state, and the surface is required as in \cite{Lewkowycz:2013nqa} to satisfy equations of motion that force it to become extremal when the conical deficit vanishes\footnote{The analysis of \cite{Lewkowycz:2013nqa} was perturbative in the conical deficit, but the issue will be discussed further in \cite{DMtoappear}.}.  While the conical angle must be constant along the brane, its value is not constrained.  Instead, one adjusts the deficit/excess to attain the desired $A(\gamma_{\bar R})$.
\end{enumerate}

However, considering a further ingredient elucidates the role played by fixed-area states.  As shown in \cite{Akers:2018fow,Dong:2018seb} the density matrix $\rho_{\bar R}$ on $\bar R$ defined by each near-eigenstate $|\psi; A \rangle$ has a flat spectrum of eigenvalues at leading order in $G$, meaning that to this order all non-zero eigenvalues of $\rho_{\bar R}$ are equal, and thus equal to $e^{-S}$ where $S$ is the von-Neumann entropy of $\rho_R$.  As a result, in a Schmidt decomposition of $|\psi; A\rangle$ with respect to ${\cal H}_R$ and ${\cal H}_{\bar R}$, one finds all Schmidt coefficients to be equal in magnitude and we have

\begin{equation}
\label{eq:RbarR}
|\psi; A\rangle = e^{-S/2}\sum_k  e^{i\theta_k} |k\rangle_R |k\rangle_{\bar R}
\end{equation}
for real angles $\theta_k$, bases $\{ |k\rangle_R\}$, $\{ |k\rangle_{\bar R} \}$ of ${\cal H}_R, {\cal H}_{\bar R}$,  and $S = A/4G +O(1)$.  So given two states $|\psi_1;A\rangle$, $|\psi_2;A\rangle$ of the form \eqref{eq:RbarR} that are CPT-conjugate on $R$ (i.e., for which there are Schmidt bases $\{ |k1\rangle_R\}$, $\{ |k2\rangle_R\}$ with $|k1\rangle_R  = (CPT)_R |k2\rangle_R$), applying the sewing-over-$R$ construction described in the introduction yields
\begin{equation}
\label{eq:RbarRsum}
|\psi_1 \# \psi_2; A\rangle = e^{-S}\sum_k  e^{i\left(\theta_{k1} + \theta_{k1} \right)} |k1\rangle_{\bar R} |k2\rangle_{\bar R}.
\end{equation}
Note that, although the final result is no longer normalized, we again find all terms to have equal weight.

In contrast, each full state $|\psi_1\rangle$, $|\psi_2\rangle$ (obtained without first fixing the area) is generally a sum over states of the form \eqref{eq:RbarR} weighted by positive functions\footnote{Phases can be absorbed into the definitions of $|\psi_1;A\rangle$, $|\psi_2;A\rangle$.} $\sqrt{P_1(A)}, \sqrt{P_2(A)}$, where $P_1(A)=P_2(A)=P(A)$ are the probabilities to find HRT-areas $A$ in each state and the probabilities must agree if the states are CPT-conjugate on $R$. The original states were thus sums over $e^{S(A)} \approx e^{A/4G}$ Schmidt terms for each $A$, with each term weighted by $e^{-S(A)/2}\sqrt{P(A)}$.
And since distinct area eigenvalues yield orthogonal states,  the sewn state $|\psi_1 \#_R \psi_2\rangle$ is then a sum of $e^{S(A)}$ Schmidt terms for each possible area $A$, with each term weighted by $e^{-S(A)}P(A)$.    As a result, the probabilities to find each $A$ in the sewn state
$|\psi_1 \#_R \psi_2\rangle$ are given by a new distribution $\tilde P(A) = N e^{-S(A)} [P(A)]^2$ for an appropriate normalization constant $N$, and the relative probabilities for different values of $A(\gamma_{\bar R})$ generally differ markedly from those in either $|\psi_1\rangle$ or $|\psi_2\rangle$.  This makes it clear that $|\psi_1 \#_R \psi_2\rangle$ cannot be described by a simple cut-and-paste construction based on the bulk geometries dual to
$|\psi_1$ and $|\psi_2\rangle$, as in particular such a construction would predict the wrong expectation value for $A(\gamma_{\bar R})$.

As described in \cite{Akers:2018fow,Dong:2018seb}, such effects should be interpreted as meaning that when one computes $|\psi_1 \#_R \psi_2\rangle$ the dominant bulk geometry shifts away from those that dominate $|\psi_1\rangle$ and $|\psi_2\rangle$.  Preventing this shift would require re-weighting the Schmidt coefficients by hand.  This approach will be discussed further in section \ref{disc}, and can be used to construct the canonical purification of \cite{Dutta:2019gen}.

On the other hand, for fixed area states generic terms appearing on the right-hand-side of \eqref{eq:RbarR} and \eqref{eq:RbarRsum} are expected to be macroscopically indistinguishable in the bulk.  In this case, \eqref{eq:RbarR} and \eqref{eq:RbarRsum} are each dominated by a single classical bulk geometry, and the above formulae certainly suggest that the geometry dual to $|\psi_1 \#_R \psi_2\rangle$ can be built by pasting together the $\bar R$ entanglement wedges from the geometries dual to the original states $|\psi_1\rangle$ and $|\psi_2\rangle$.

\section{Sewing, Cutting, and Pasting in fixed-area states}
\label{arg}

We now proceed to the main argument. For simplicity, we consider states invariant under CPT and use Euclidean path integrals.  But -- at least at this order in $G$ -- the more general case is identical with the understanding that Euclidean path integrals are replaced by Schwinger-Keldysh path integrals as in \cite{Dong:2016hjy} and also that the bulk gluing of entanglement wedges becomes the construction of \cite{Engelhardt:2018kcs}.  Although \cite{Engelhardt:2018kcs} focussed on gluing a single entanglement wedge to a CPT-copy of itself, it also discussed more general gluings and found matching conditions that are automatically satisfied in our context\footnote{\label{foot:sew}In more detail, they found that a wedge can always be glued to its CPT-conjugate.  Since by assumption the entanglement wedges $W_{R_1}$, $W_{R_2}$ of $R$ in the two original geometries are CPT-conjugate, we could thus have glued them together.   But the two original geometries were clearly well-defined, so the pairs of wedges $(W_{R_1}, W_{{\bar R}_1})$, $(W_{R_2}, W_{{\bar R}_2})$ also satisfy the matching conditions.  Thus $W_{{\bar R}_1}$ must have the same relevant data at $\gamma_{\bar R}$ as $W_{{R}_2}$, and so can be matched to $W_{{\bar R}_2}$ to yield a spacetime at least as smooth as the roughest of those associated with $(W_{R_1}, W_{{\bar R}_1})$, $(W_{R_1},W_{R_2})$, or $(W_{R_2}, W_{{\bar R}_2})$.  In particular, for smooth original geometries, smoothness of the canonical purification \cite{Dutta:2019gen} defined by gluing $W_{R_1}$ to $W_{R_2}$ implies that our final geometry is also smooth and prohibits the possibility mentioned in the introduction of a mild shock emanating from $\gamma_{\bar R}$ describing a discontinuity, but not a divergence, in the shear or matter stress tensor.}.  In the general case we must take the states $|\psi_1\rangle, |\psi_2\rangle$ to be associated with CFT spacetimes $M_1,M_2$ containing CPT-conjugate regions $R_1,R_2$ over which the states will be sewn together to define $|\psi_1\rangle, |\psi_2\rangle$.  But in the time-symmetric context the regions become isometric in a time-orientation-preserving sense and it is natural to write $R_1 = R_2 = R$ while still requiring any sources on $R_1$ to be CPT-conjugate to those on $R_2$.

As mentioned above, we suppose the states $|\psi_1\rangle, \psi_2\rangle$ to be prepared by Euclidean path integrals.  Since we assumed the states are CPT-invariant, we may take each path integral to be invariant as well.  As a result, the CFT spacetime of each path integral has a preferred surface left fixed by the action of CPT.  We refer to this surface as $t=0$. If each state is described by a unique semi-classical geometry in the bulk, these geometries must have similarly-defined surfaces that we may again call $t=0$.

\begin{figure}[h]
\centering
\includegraphics[width =0.35\textwidth]{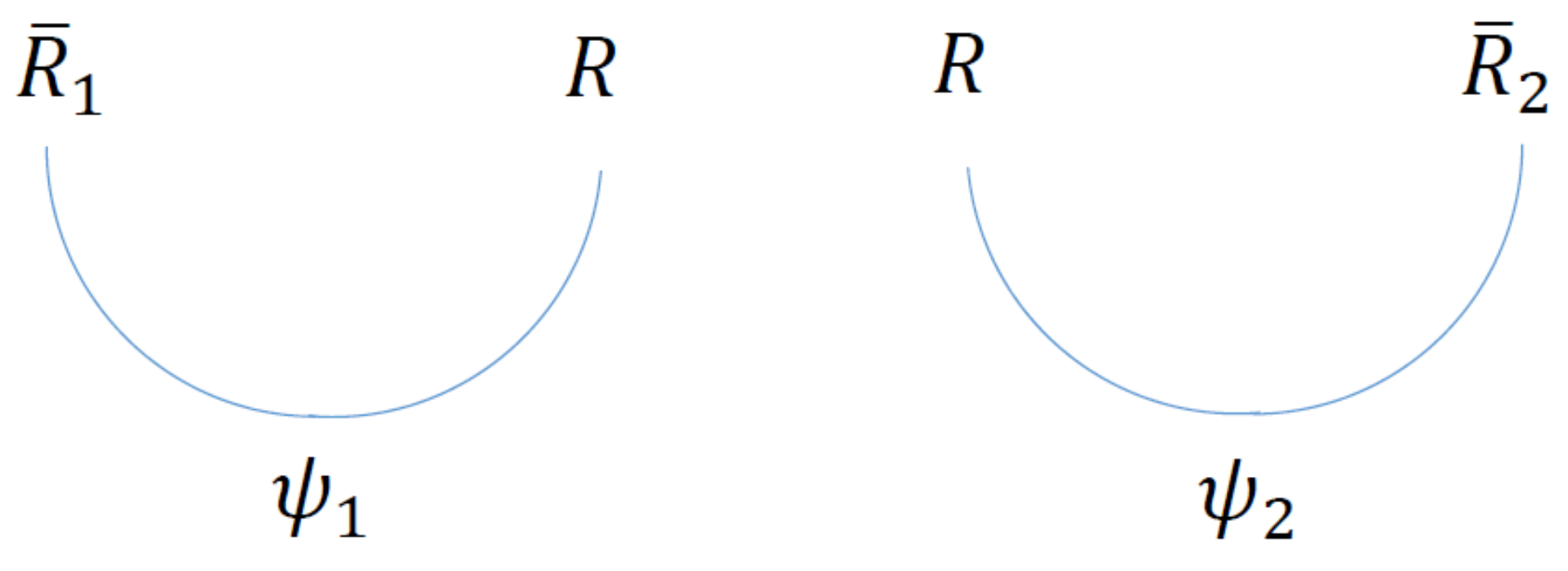}\ \ \ \ \includegraphics[width =0.35\textwidth]{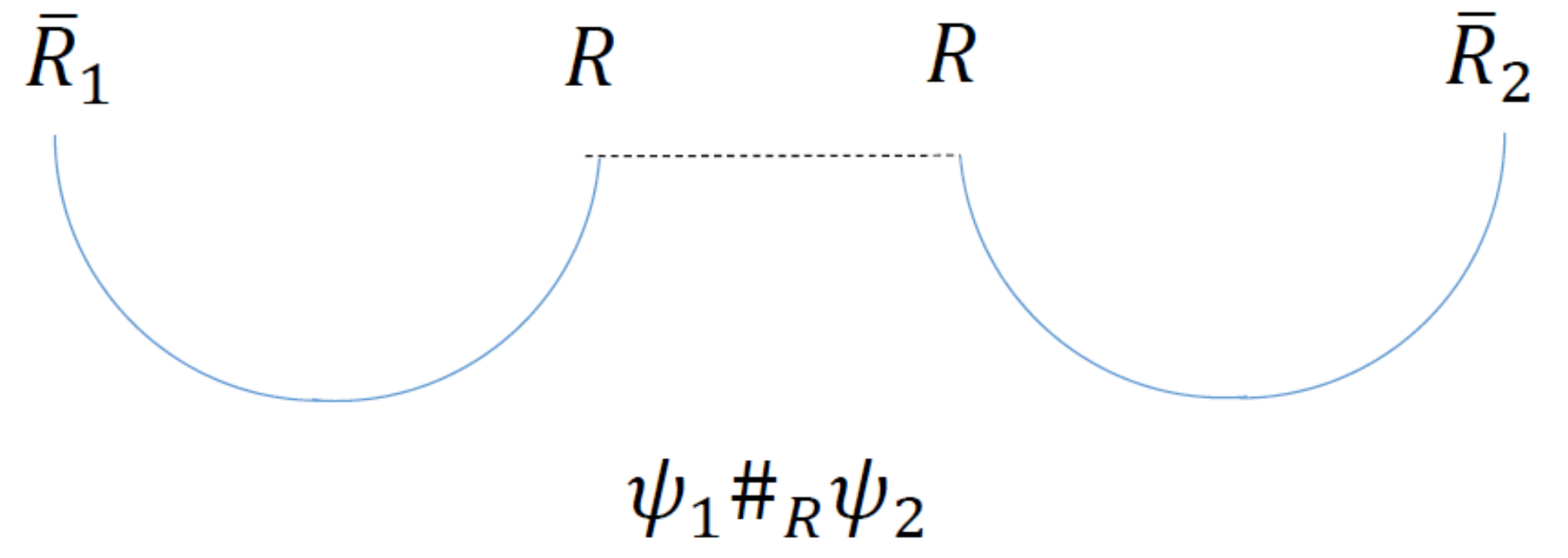}\\ \medskip \medskip \medskip
\includegraphics[width =0.35\textwidth]{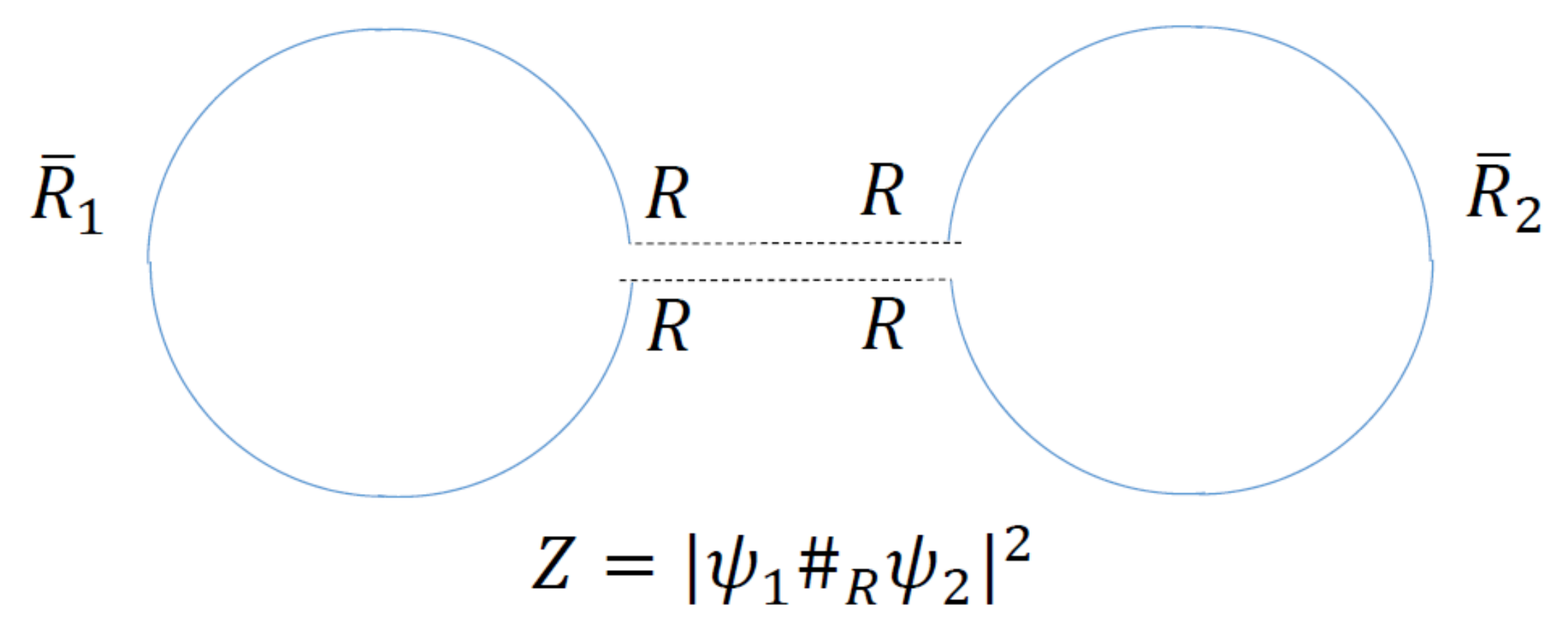}\includegraphics[width =0.35\textwidth]{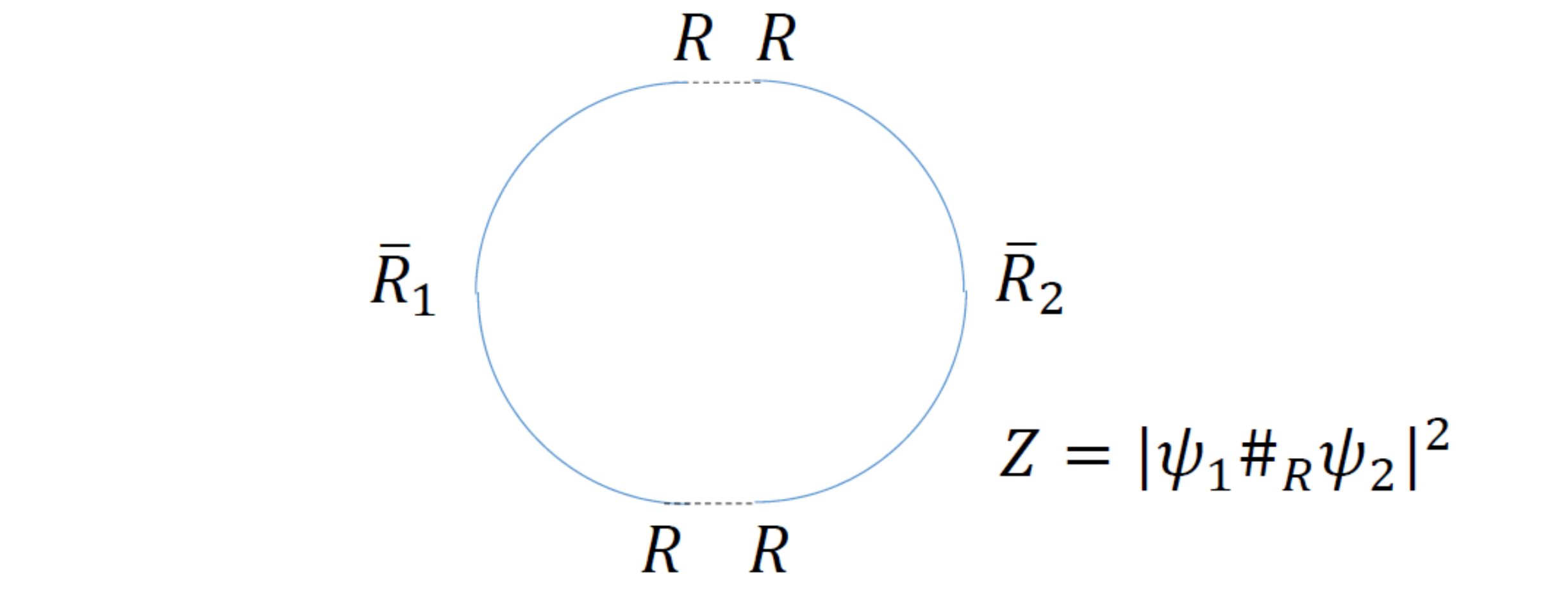}\\ \medskip \medskip \medskip
\includegraphics[width =0.35\textwidth]{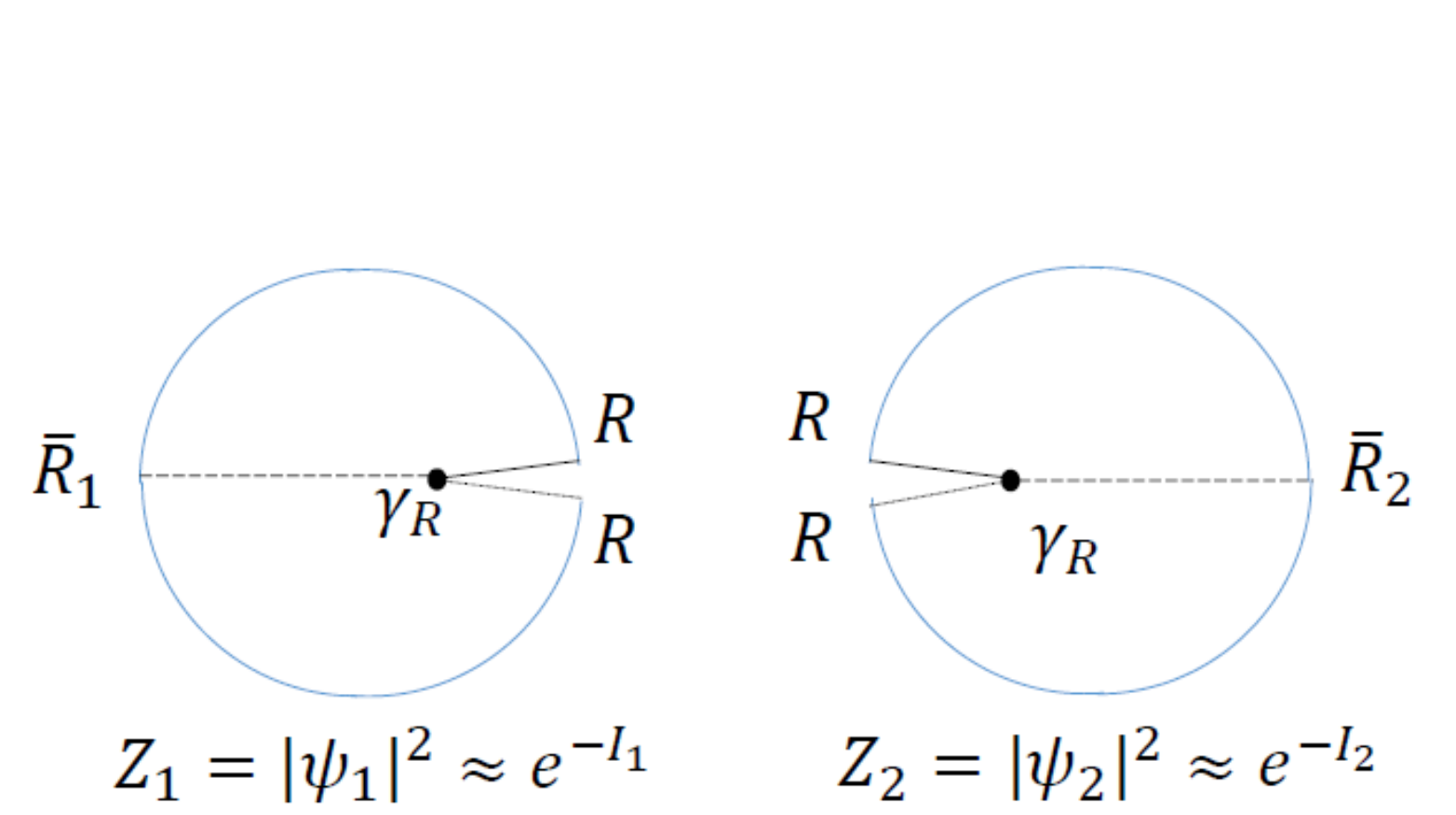} \includegraphics[width =0.35
\textwidth]{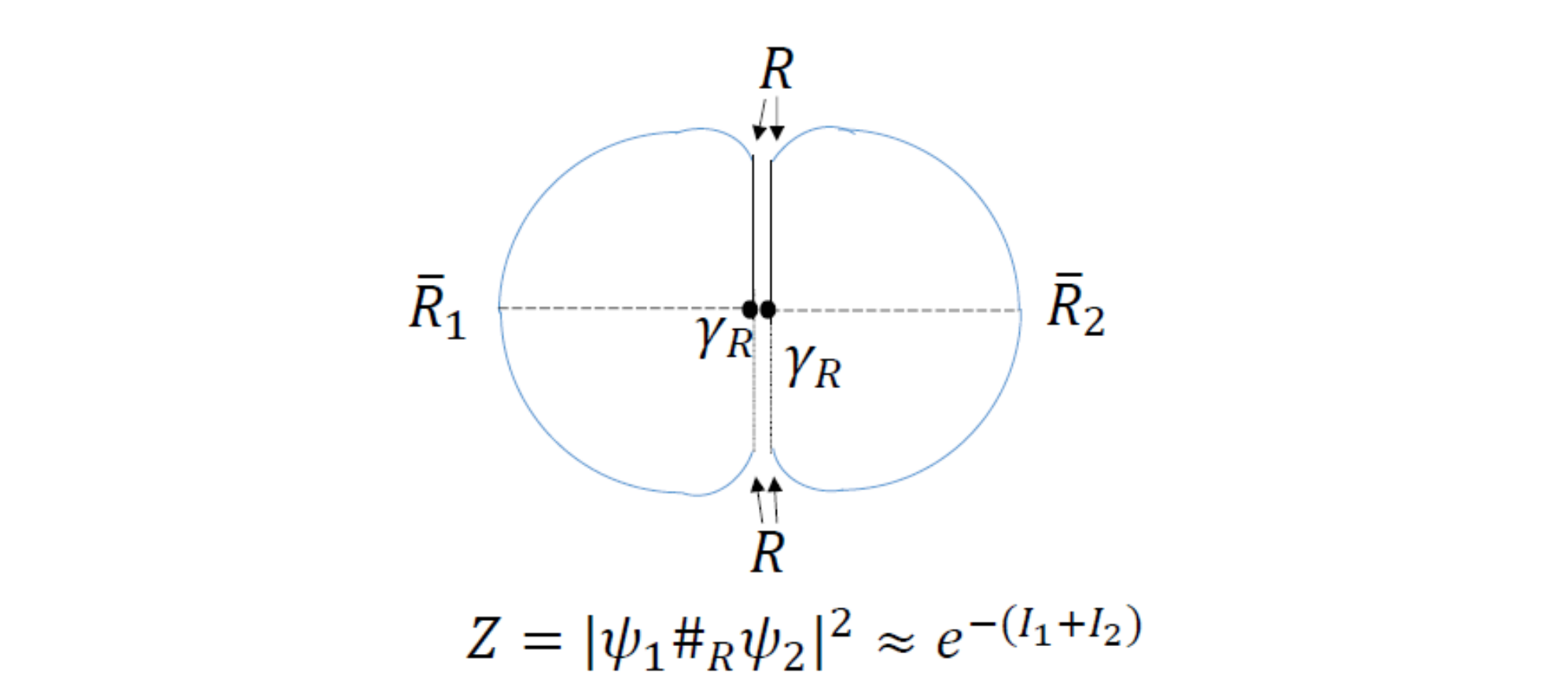}\\
\caption{Constructing a bulk saddle for $Z = |\psi_1 \#_R \psi_2|^2$.  {\bf Top Row:} When the regions $R=R_1$ and $R= R_2$ are CPT-conjugate,  path integrals  (semi-circles, shown separately at left) for pure CFT states $\psi_1$ on $R_1{\bar R}_1$ and $\psi_2$ on $R_2{\bar R}_2$ can be sewn together to give a path integral representation of $\psi_1 \#_R \psi_2$ on ${\bar R_1}{\bar R}_2$ (right). In the time-symmetric case, the path integrals may be taken to be Euclidean.  {\bf Center Row:} The corresponding path integral for the norm $Z = |\psi_1 \#_R \psi_2|^2$ of the sewn state is shown at left. Another representation of the same path integral is shown at right, differing only by a diffeomorphism that compresses to semi-circles those path-integral-parts that were (nearly) entire circles at left.   {\bf Bottom Row:}  The dominant bulk saddles for path integrals computing the individual norms $Z_1 = |\psi_1|^2$, $Z_2 = |\psi_2|^2$ can be cut open through the entanglement wedges of $R_1, R_2$ as shown at left.  Since the two states are CPT conjugate on $R$, the data on the upper cuts is also CPT-conjugate, as is the data on the lower cuts.  As a result, we can join the cut saddles to build a saddle for $Z = |\psi_1 \#_R \psi_2|^2$ as shown at right. Dashed lines indicate $t=0$ in the wedges dual to ${\bar R}_1$, ${\bar R}_2$. The $t=0$ surface in the glued saddle (right) is obtained by gluing together these parts of $t=0$ from the original saddles (left).}
\label{fig:contractPIs}
\end{figure}

As our construction is local in time, it suffices for the region $R$ to be a partial Cauchy surfaces, perhaps slightly thickened in time to remove UV divergences.  Sewing our states together along $R$ to build the new state\footnote{In the general context without time-symmetry, one should use the more cumbersome notation $|\psi_1 \#_{R_1,R_2} \psi_2 \rangle$.} $|\psi_1 \#_R \psi_2 \rangle$ is then merely a matter of stitching together the path integrals as shown in the top row of figure \ref{fig:contractPIs}.  The sewn CFT state lives on a new spacetime in which regions $\bar R_1$, $\bar R_2$ now partition a CFT cauchy surface.  So in any new bulk the HRT-surface for $\bar R_1$ will also be the HRT-surface for $\bar R_2$.  For simplicity, we simply refer to it as $\gamma_{\bar R}$.

We wish to consider the state $|\psi_1 \#_R \psi_2; A \rangle$ defined by projecting
$|\psi_1 \#_R \psi_2 \rangle$ onto a small range of eigenvalues for the HRT-area $A(\gamma_{\bar R})$ centered on the eigenvalue $A$.  Since the associated projection operator $\Pi_A$ satisfies $\Pi_A^2 =\Pi_A$, and since $\Pi_A$ can be reconstructed on both $\bar R_1$ and $\bar R_2$ in the original states, projecting first to yield $|\psi_1; A \rangle$, $|\psi_2; A \rangle$ and then sewing these states together on $R$ must yield the same state $|\psi_1 \#_R \psi_2; A \rangle$; see \cite{Dong:2018seb} for further discussion of such issues.

Using the sew-first, project-second construction of $|\psi_1 \#_R \psi_2; A \rangle$, points (2) and (3) from section \ref{review} imply that the norm
\begin{equation}
Z = \langle \psi_1 \#_R \psi_2; A | \psi_1 \#_R \psi_2; A \rangle
\end{equation}
is given by sewing together the partition-function path integrals giving separately the norms $\langle \psi_1 |\psi_1 \rangle$, $\langle \psi_2 |\psi_2 \rangle$, and allowing the bulk dual to contain a conical singularity anchored to $\partial \bar R_1 = \partial \bar R_2$.  The value of the conical deficit/excess is to be turned to yield the desired value for $A(\gamma_{\bar R})$.

Furthermore, the two fixed-area states $|\psi_1; A \rangle, |\psi_2;A \rangle$ are associated with similar bulk path integrals fixing the areas of $\gamma_{\bar R_1}$ and $\gamma_{\bar R_1}$  to the same desired value.  As noted above, the leading bulk saddles $g_1(A),g_2(A)$ for each will have a well-defined $t=0$ surface.  And each $t=0$ surface will intersect the relevant entanglement wedge of $R$.

We now construct a saddle $g(A) = g_1(A) \#_R g_2(A)$ for $Z$ by the following procedure.  First, cut open the Euclidean saddles $g_1(A), g_2(A)$ along the parts of the $t=0$ surfaces that lie in the entanglement wedges of $R$.  Second, glue together the resulting Euclidean geometries by gluing together the upper parts of the two cuts and by also gluing together the lower parts of the two cuts; see figure \ref{fig:contractPIs} (right).  This introduces a conical excess of $2\pi$ along $\gamma_{\bar R}$, but that is allowed in a fixed-area state.  Indeed, we are instructed to tune a conical singularity there to attain the desired area $A$.

Since $g_1(A), g_2(A)$ already had the correct value of this area, and since continuity of $g_1(A), g_2(A)$ allows us to take a limit in which $\gamma_{\bar R}$ is apporached from the CPT-conjugate entanglement wedges of $R$, we see that both $g_1(A)$ and $g_2(A)$ induce the same geometry on $\gamma_{\bar R}$. This is also the geometry induced on $\gamma_{\bar R}$ by the glued saddle $g(A) = g_1(A) \#_R g_2(A)$.  It follows that the area constraint is satisfied by $g(A) = g_1(A) \#_R g_2(A)$.  The geometry $g(A) = g_1(A) \#_R g_2(A)$ also clearly satisfies all equations of motion away from $\gamma_{\bar R}$ and all desired boundary conditions at AdS-infinity.  To see that the equations of motion are satisfied on the gluing surface away from $\gamma_{\bar R}$, recall that  the saddles $g_1(A), g_2(A)$ are CPT-conjugate in the entanglement wedge of $R$.  Since the cuts lie in that wedge, the saddles $g_1(A), g_2(A)$ induce CPT-conjugate data on the codimension-1 gluing surface. This means that $g(A) = g_1(A) \#_R g_2(A)$ is smooth across this surface away from $\gamma_{\bar R}$, and thus that the desired equations of motion are satisfied.   We conclude that $g(A) = g_1(A) \#_R g_2(A)$ yields a valid saddle for the bulk path integral computing $Z$.

Assuming that this saddle dominates the path integral, the state $| \psi_1 \#_R \psi_2; A \rangle$ must be dual to a Lorentz-signature bulk gometry given by evolving initial data from the $t=0$ surface of $g(A) = g_1(A) \#_R g_2(A)$.  By this we again mean the surface invariant under the CPT-symmetry of the path integral.  But as can be seen from the last line of figure \ref{fig:contractPIs}, in $g(A)= g_1(A) \#_R g_2(A)$ this bulk $t=0$ surface is given by sewing together the parts of the $t=0$ surfaces of $g_1(A)$ and $g_2(A)$ lying in the entanglement wedges of ${\bar R}_1, {\bar R}_2$.  This is precisely the desired result:  For fixed-area states, the bulk dual of the natural CPT-sewing of the CFT states over $R$ is defined by gluing together the associated bulk entanglement wedges for the complementary regions $\bar R_1$ and $\bar R_2$.

\section{Discussion}
\label{disc}

The above work studied the natural CPT-sewing of two quantum field theory states over CPT-conjugate regions $R_1,R_2$ for holographic CFTs dual at leading order in the bulk Newton constant $G$.   In the time-symmetric case, $R_1$ and $R_2$ are isometric in a time-orientation-preserving sense and we can write $R_1 = R_2 = R$ with the understanding that $R$ contains CPT-conjugate sources in the two CFT spacetimes.   We considered states  $|\psi_1; A\rangle$, $|\psi_2; A\rangle$ of fixed bulk HRT-area $A$ that are also CPT-conjugate to each other in $R$, and argued that the bulk geometry dual to the sewn state $|\psi_1 \#_R \psi_2 ; A\rangle$ can be obtained from the geometries $g_1(A)$, $g_2(A)$ dual to the original states $|\psi_1; A\rangle$, $|\psi_2; A\rangle$ by extracting from $g_1(A), g_2(A)$ the entanglement wedges of the regions $\bar R_1, \bar R_2$ complementary to $R$ and gluing these wedges together to define $g(A) = g_1(A) \#_R g_2(A)$ as in the last line of figure \ref{fig:contractTNs}.    The work above assumed the bulk to be described by Einstein-Hilbert gravity, but using results from the forthcoming work \cite{DMtoappear} and assuming extensions of the matching conditions in \cite{Engelhardt:2018kcs} to the higher derivative context, analogous conclusions will continue to hold with arbitrary perturbative higher-derivative corrections.

As discussed in detail in footnote \ref{foot:sew}, 
the wedges of ${\bar R}_1, {\bar R}_2$ will always satisfy the matching conditions of \cite{Engelhardt:2018kcs}.  This means that the spacetime $g(A) = g_1(A) \#_R g_2(A)$ is well-defined, and in particular satisfies the gravitational constraint equations at $\gamma_{\bar R}$.  However, the fields need not necessarily be as smooth as in the original $g_1(A), g_2(A)$.  While continuity of fields at $\gamma_{\bar R}$ follows from taking limits from the $R_1,R_2$ entanglement wedges, normal derivatives at $\gamma_{\bar R}$ typically change by a sign across this surface.  So unless such normal derivatives vanish, $g(A) = g_1(A) \#_R g_2(A)$ will contain a mild shockwave propagating along both the past- and future-directed null congruences orthogonal to $\gamma_{\bar R}$ describing a discontinuity, but not a divergence, in the null congruence shear and the matter stress tensor.  The same phenomonon occurs in the bulk geometry $g_{\sqrt{\rho_R}}$ dual to the canonical purification $\sqrt{\rho_R}$ (see \cite{Dutta:2019gen}), and in fact our $g(A) = g_1(A) \#_R g_2(A)$ is at least as smooth as the roughest of $g_1(A), g_2(A), \sqrt{\rho_R}$.  It would be interesting to understand better how such shocks are encoded in the CFT states, presumeably by studying the modular zero modes discussed in \cite{Faulkner:2017vdd}.

Our argument assumed the bulk path integral for $|\psi_1 \#_R \psi_2 ; A\rangle$  to be dominated by a natural saddle.  This is directly analogous to similar assumptions in \cite{Dong:2018seb} and \cite{Dutta:2019gen}.  A related assumption was also made in the classic work by Lewkowycz and Maldacena \cite{Lewkowycz:2013nqa}, though there the relevant saddle was required to dominate only in the limit $n\rightarrow 1$ of trivial replica number.  As in the works above, this assumption leads to an elegant picture.  But as always it deserves further investigation.

In fact, for our duality to hold the states   $|\psi_1; A\rangle$, $|\psi_2; A\rangle$ need only be CPT-conjugate on $R$ to leading order in the bulk Newton constant $G$.  That is because our construction required only that the states have bulk duals in which the $R$ entanglement wedges are CPT-conjugate, and also because we focused only on leading-order results in $G$.  Now, as usual, one may also consider higher-order corrections.  But the leading-order construction implies the bulk dual of
$|\psi_1 \#_R \psi_2 ; A\rangle$  to have quantum fields on  $g_1(A) \#_R g_2(A)$ defined by the path integral over the above-mentioned saddle even when the states on $R$ are fail to be CPT-conjugate to higher orders, and this path integral in fact simply implements the natural CPT-sewing of bulk quantum states.  We thus see that a consistent picture of the bulk and its higher-order corrections can be obtained even when the original states are only CPT-conjugate on $R$ at leading order in $G$.

When the states $|\psi_1; A\rangle$, $|\psi_2; A\rangle$ are CPT-conjugate to this order everywhere, our $|\psi_1 \#_R \psi_2 ; A\rangle$ gives the canonical purification described in \cite{Dutta:2019gen} of the density matrix $\rho_{\bar R_1}$ defined by $|\psi_1; A\rangle$.  This state is also the canonical purification of the density matrix $\rho_{\bar R_2}$ defined by $|\psi_2; A\rangle$.   However, the canonical purification is also defined for states $|\psi_1\rangle, |\psi_2\rangle$ in which the area is not fixed.  In general, one may think of this purification as $\sqrt{\rho_{{\bar R}_1}}$ (or as $\sqrt{\rho_{{\bar R}_2}}$) defined by the positive operator-square-root of $\rho_{{\bar R}_1}  (\rho_{{\bar R}_2}$) and with such operators reinterpreted as states on a doubled Hilbert space by combining CPT with the adjoint operation to linearly map bra-vectors into ket-vectors via $(CPT_{\bar R_1} \alpha |\psi\rangle)^\dagger = \alpha \langle \psi | (CPT_{\bar R_1})^\dagger$.

From the perspective of fixed-area states, the point of using $\sqrt{\rho_{{\bar R}_1}}$, $\sqrt{\rho_{{\bar R}_2}}$ instead of simply ${\rho_{{\bar R}_1}}$, ${\rho_{{\bar R}_2}}$ is that -- when interpreted as states on the doubled Hilbert space -- the former have Schmidt coefficients $c_k$ of the same magnitudes as in the original states  $|\psi_1\rangle, |\psi_2\rangle$.  In contrast, the Schmidt coefficients defined by ${\rho_{{\bar R}_1}}$, ${\rho_{{\bar R}_2}}$  are $|c_k|^2$.  As described in section \ref{review}, this means that (when interpreted as pure states on the doubled Hilbert space) bulk duals to $\sqrt{\rho_{{\bar R}_1}}$, $\sqrt{\rho_{{\bar R}_2}}$ will have the same area for the HRT surface as those dual to  $|\psi_1\rangle, |\psi_2\rangle$. In contrast, one generally finds a very different HRT-area in any bulk dual to this pure-state-on-a-doubled-Hilbert-space interpretation of ${\rho_{{\bar R}_1}}$, ${\rho_{{\bar R}_2}}$.

In this way, the natural CPT-sewing tends to change the weighting of bulk geometries associated with different components of CFT states $|\psi_1\rangle, |\psi_2\rangle$.  This can be counteracted by re-weighting the various components by hand, or equivalently by using a modified sewing operation.  Since we require $|\psi_1\rangle, |\psi_2\rangle$ to be CPT-conjugate on $R$, they define CPT-conjugate density matrices ${\rho_{{R}_1}}$, ${\rho_{{R}_2}}$.  Regardless of whether these are fixed-area states, one may thus choose to sew the states together using not just the natural CPT-sewing of \eqref{eq:natural}, but instead using

\begin{equation}
\label{eq:modified}
B_{\rho_{\bar R}}(\psi_1, \psi_2) := \Bigl( (CPT_R)\psi_2, (\rho^{-1/2}_{{R}_1}\otimes {\mathds 1}_{{\bar R}_1} )\psi_1 \Bigr)
= \Bigl( (CPT_R)  (\rho^{-1/2}_{{R}_2}\otimes {\mathds 1}_{{\bar R}_2} ) \psi_2,  \psi_1 \Bigr).
\end{equation}

Here the action of 
$(\rho^{-1/2}_{{R}_i}\otimes {\mathds 1}_{{\bar R}_i} )$ is defined using the factorization ${\cal H}_{CFT} = {\cal H}_{R_i} \otimes {\cal H}_{{\bar R}_i}$ and
the subscript ${\rho_{\bar R}}$ on $B$ indicates that the sewing map depends on the pair of states on which it will act.
The latter contrasts with the universal map \eqref{eq:natural} that can be used for arbitrary fixed-area states.  It is the price to be paid to prevent the dominant bulk geometries on $\bar R_1, \bar R_2$ from shifting when we sew together two states in the CFT.  However, as a bonus when the states on $R_1, R_2$ are CPT conjugate to higher orders in $G$, using the modified sewing \eqref{eq:modified} also ensures that the higher-order bulk quantum states are unchanged by the sewing within $\bar R_1, \bar R_2$.  As in \cite{Dutta:2019gen}, new features of the bulk quantum state can then arise only to the future or past of the gluing surface $\gamma_{\bar R}$.

Let us denote the result of the above modified sewing as $|\psi_1 \#_{\rho_{\bar R}} \psi_2\rangle$, again indicating that the sewing depends on the density matrix and not just on the choice of region $R$.  When $|\psi_1\rangle$ is CPT-conjugate to $|\psi_2\rangle$ on the full CFT, our  \eqref{eq:modified} gives $|\psi_1 \#_{\rho_{\bar R}} \psi_2\rangle = \sqrt{\rho_{\bar R}}$; i.e., this is the canonical purification of \cite{Dutta:2019gen}.  But to ensure that the dominant value $A_{dom}$ of $A(\gamma_{\bar R})$ is the same in $|\psi_1 \#_{\rho_{\bar R}} \psi_2\rangle$ as in $|\psi_1\rangle$, $|\psi_2\rangle,$
it suffices for the states to be CPT-conjugate only on $R$. This ensures that $|\psi_1\rangle$ and $|\psi_2\rangle$ yield the same probabilities $P(A_{\gamma_{\bar R}})$ for each possible HRT-area $A_{\gamma_{\bar R}}$, and thus that these are also the probabilities in $|\psi_1 \#_{\rho_{\bar R}} \psi_2\rangle$.  Furthermore, at leading order in $G$ we have
\begin{equation}
|\psi_1 \#_{\rho_{\bar R}} \psi_2\rangle \approx |\psi_1 \#_{\rho_{\bar R}} \psi_2; A_{dom}\rangle
\approx |\psi_1; A_{dom}\rangle \#_{R} |\psi_2; A_{dom}\rangle,
\end{equation}
where the right-most expression denotes the natural sewing of two fixed-area states using the original map \eqref{eq:natural}.  It thus follows that $|\psi_1 \#_{\rho_{\bar R}} \psi_2\rangle $ is dual to the bulk geometry $g_1(A_{dom}) \#_R g_2(A_{dom})$ shown to be dual to $|\psi_1; A_{dom}\rangle \#_{R} |\psi_2; A_{dom}\rangle$ in section \ref{arg}.  And since $A_{dom}$ is the dominant value of $A(\gamma_{\bar R})$ in each case, the geometries $g_1(A), g_2(A)$ also describe the bulk duals to to the full original states $|\psi_1\rangle$, $|\psi_2\rangle$.  In other words, for general CFT states whose HRT-areas need not already be fixed, the bulk cut-and-paste operation shown in the final line of figure \ref{fig:contractTNs} is dual to the CFT sewing operation $|\psi_1 \#_{\rho_{\bar R}} \psi_2\rangle$.  One should also be able to give an alternative argument more directly parallel to that in \cite{Dutta:2019gen}.

Our results generalize not only the construction of \cite{Dutta:2019gen}, but also previous analyses \cite{Balasubramanian:2014hda,Marolf:2015vma} of bulk duals of CFT sewing in 2+1 multi-boundary wormholes.    In retrospect, the complications of \cite{Balasubramanian:2014hda,Marolf:2015vma} were largely due to the shift of the classical bulk saddle under the natural CPT-sewing of the CFT states when fixed-area constraints are not included.

Let us now discuss two further generalizations.  First, one may note that the results of \cite{Engelhardt:2018kcs} allow two entanglement wedges with appropriately-compatible data on the HRT-surfaces to be directly sewn together without first embedding each in a larger geometry, and certainly without requiring the complementary wedges in that geometry to be CPT-conjugate.  It is thus natural to ask if there is a good CFT dual to this more general bulk gluing.  We suspect that the answer is affirmative, but also that that constructing the dual state will require significant further input.  Again, the fixed-area-state analysis provides some insight.  The bulk geometries should have the same induced metric on the HRT surface, and in particular should have the same HRT-area.  For fixed-area states, to leading order in $G$ the dual density matrices $\rho_1, \rho_2$  should then both be proportional to projections onto subspaces of dimension $e^{S}$ which we call ${\cal H}_1, {\cal H}_2$.  Choosing a pure state on the product system whose reduced density matrices are $\rho_1, \rho_2$ is then equivalent to choosing a unitary map from the dual Hilbert space ${\cal H}_1^*$ to ${\cal H}_2$.  But since ${\cal H}_1^*$ and ${\cal H}_2$ are large, there is a correspondingly large space of such unitary maps.  In contrast, we expect that only a small subset will give CFT states with semi-classical bulk duals.  In particular, in the bulk quantum fields should approach the vacuum in the deep UV near the HRT-surface, so using the bulk-to-boundary dictionary from \cite{Faulkner:2017vdd} this should impose requirements on the CFT state associated with certain modular zero modes.

Turning now to a second generalization, let us return to the context where we start with a full pure CFT state $|\psi\rangle$.  Given a single such state, one may be interested in sewing $|\psi \rangle$ to itself in the following sense:  Suppose that there are spacelike-separated regions $R_1, R_2$ in the CFT spacetime for which the metric and sources on $R_1$ are CPT-conjugate to those on $R_2$, and for which the density matrices $\rho_{R_1}$, $\rho_{R_2}$ defined by $|\psi \rangle$ are also CPT-conjugate.  Then one might also consider taking the dual bulk geometry $g_0$ and excising both the wedge dual to $R_1$ {\it and} the wedge dual to $R_2$.  If $I(R_1,R_2)$ vanishes to leading order in $1/G$, this leaves only the wedge dual to $\overline{R_1 \cup R_2} = {\bar R}_1 \cap {\bar R}_2$.  One might then wish to glue the two edges $\gamma_{\bar {R_1}}$, $\gamma_{\bar {R_2}}$ of this wedge to each other and then ask if there is a good CFT dual.

Our construction can indeed be used to give such a CFT dual when the wedges of $R_1,R_2$ are sufficiently well separated in the bulk.  In particular, we show below that this is the case when there is an
intermediate CFT region $R_{int}$ i) satisfying ${\bar R}_2 \supset R_{int} \supset R_1$, ii) having no leading-order mutual information with $R_2$ (so that $I(R_{int}, R_2) = O(1)$, and iii) such that ${\bar R}_{int}$ has no leading-order mutual information with $R_1$ (so that $I({\bar R}_{int}, R_1) = O(1)$).

Under such conditions, the desired sewing can be accomplished by instead sewing the original state $|\psi\rangle$ to the canonical purification $\sqrt{\rho_{R_{int}}}$ of $\rho_{R_{int}}$, where the sewing takes place along a new region $R_{sew}$.  In $|\psi \rangle$, we take $R_{sew} = R_{int} \cup R_2$, while in $\sqrt{\rho_{R_{int}}}$ we take we take $R_{sew} = {\bar R}_{int} \cup R_1$. To apply our earlier arguments we must then show the corresponding bulk wedges to be CPT-conjugate.  But condition (ii) above means that in $|\psi\rangle$ the entanglement wedge of $R_{sew}$ is the unions of the wedges of ${R}_{int}$ and of $R_2$; see top two lines of figure \ref{fig:SelfGlue} below.  And since the geometry dual to $\sqrt{\rho_{R_{int}}}$ agrees with that dual to $|\psi \rangle$ in the entanglement wedge of $R_{int}$, condition (ii) similarly implies that the entanglement wedge of ${R}_{sew}$ in $\sqrt{\rho_{R_{int}}}$ is the union of the wedges of  ${\bar R}_{int}$ and of $R_1$.  Since the wedges of $(R_{int}, {\bar R}_{int})$ and $(R_2, R_1)$ form CPT-conjugate pairs, the entanglement wedges of $R_{sew}$ in the two spacetimes are then also CPT-conjugate.

It remains only to compute the bulk geometry dual to the properly-sewn state.  This geometry clearly consists of the wedge dual to $\bar R_{sew} = R_{int}\setminus R_1$ from $\sqrt{\rho_{R_{int}}}$ glued to the wedge dual to $\bar R_{sew} = {\bar R}_{int}\setminus R_2$ from $|\psi\rangle$.  But condition (ii) above means that the second wedge is given by removing the entanglement wedge of $R_2$ from the wedge of ${\bar R}_{int}$. And since the geometry of the wedge dual to $R_1$ is identical in both $|\psi\rangle$ and $\sqrt{R_{int}}$,  condition (iii) similarly implies that the first wedge is given by removing the entanglement wedge of $R_1$ from the wedge of $R_{int}$.  Gluing these together along $\gamma_{R_{sew}}$ then gives the entanglement wedge of $R_{int} \cup {\bar R}_{int}$ (i.e., the full bulk spacetime dual to the original state $|\psi \rangle$), with the wedges of $R_1$ and $R_2$ removed and with $\gamma_{R_1}$ glued

\begin{figure}[H]
\centering
\includegraphics[width =0.35\textwidth]{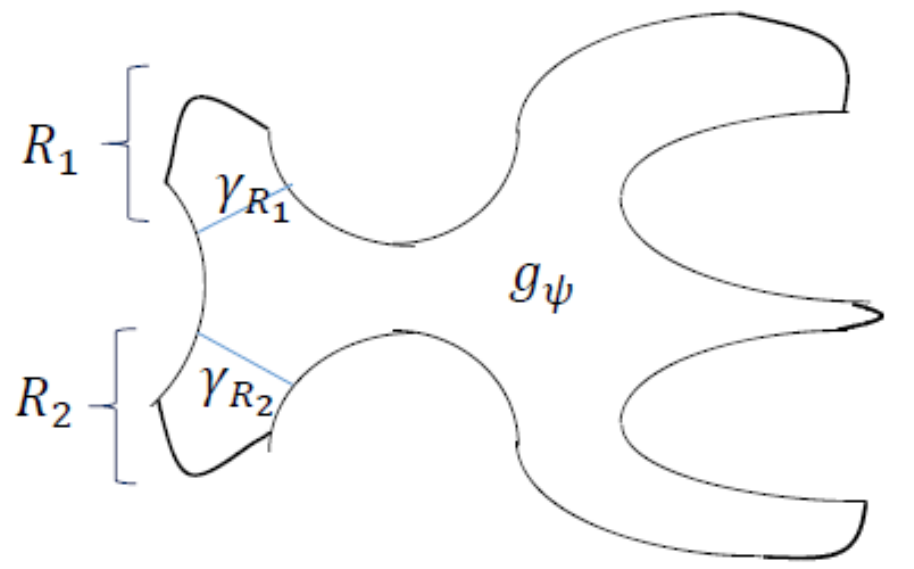}\includegraphics[width =0.35\textwidth]{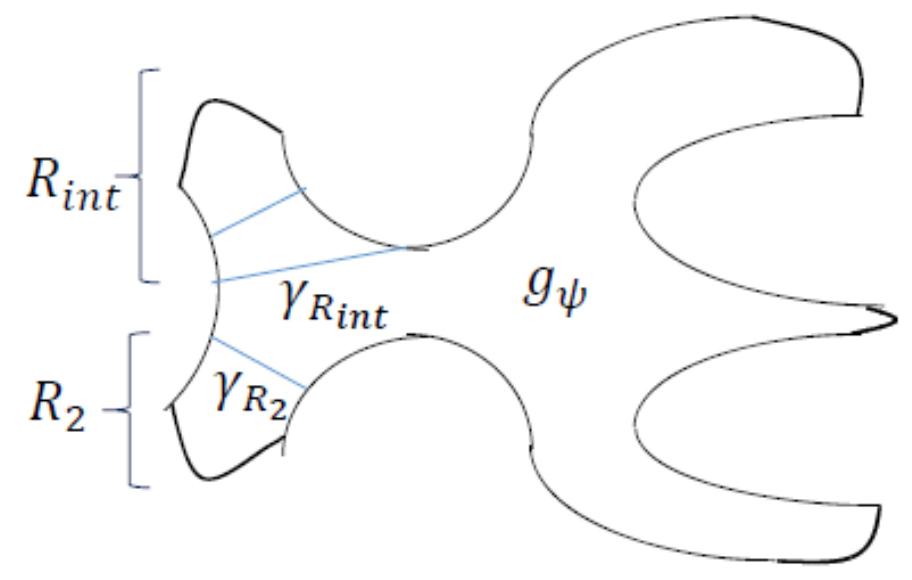}\\
\includegraphics[width =0.35\textwidth]{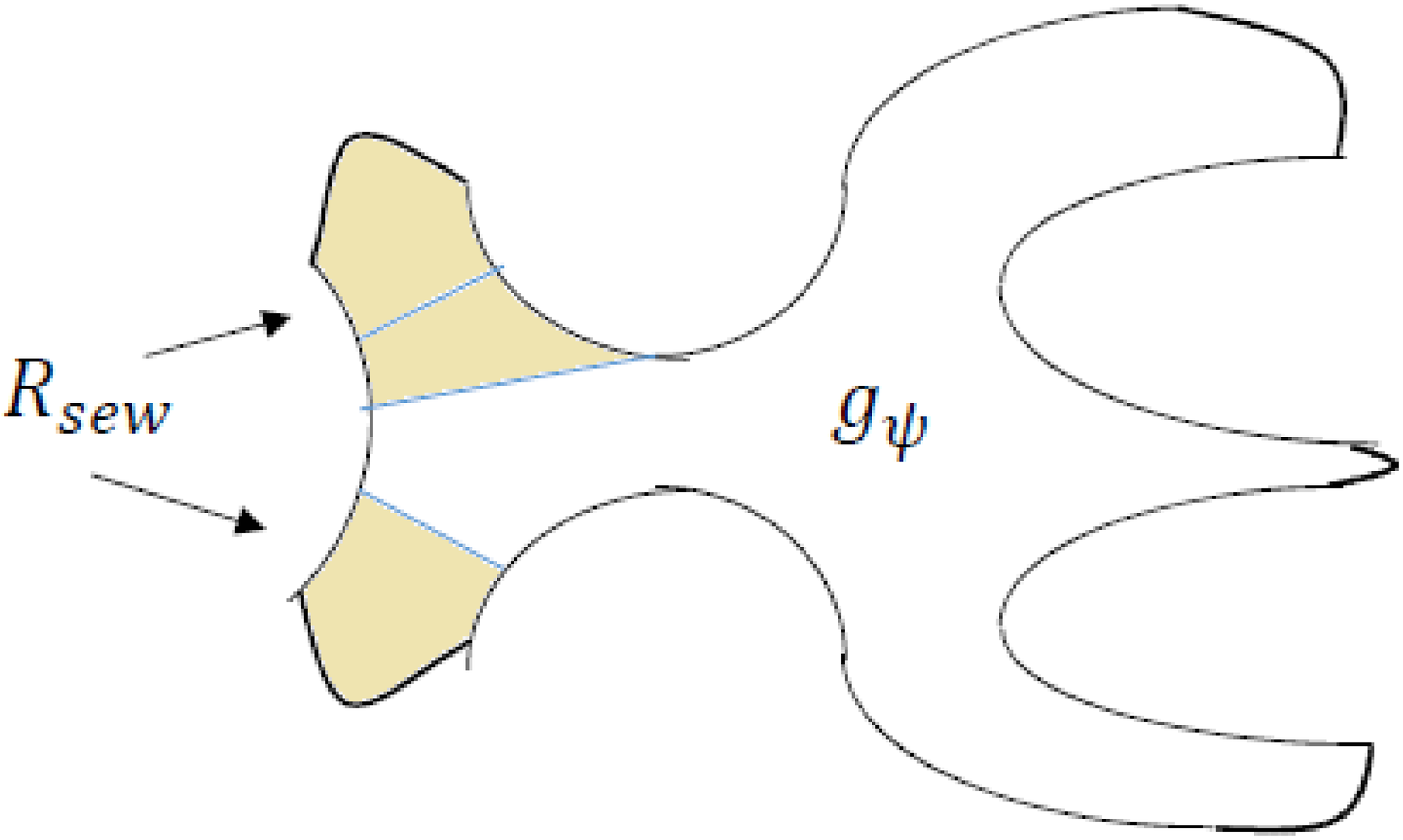}\ \ \ \ \includegraphics[width =0.35\textwidth]{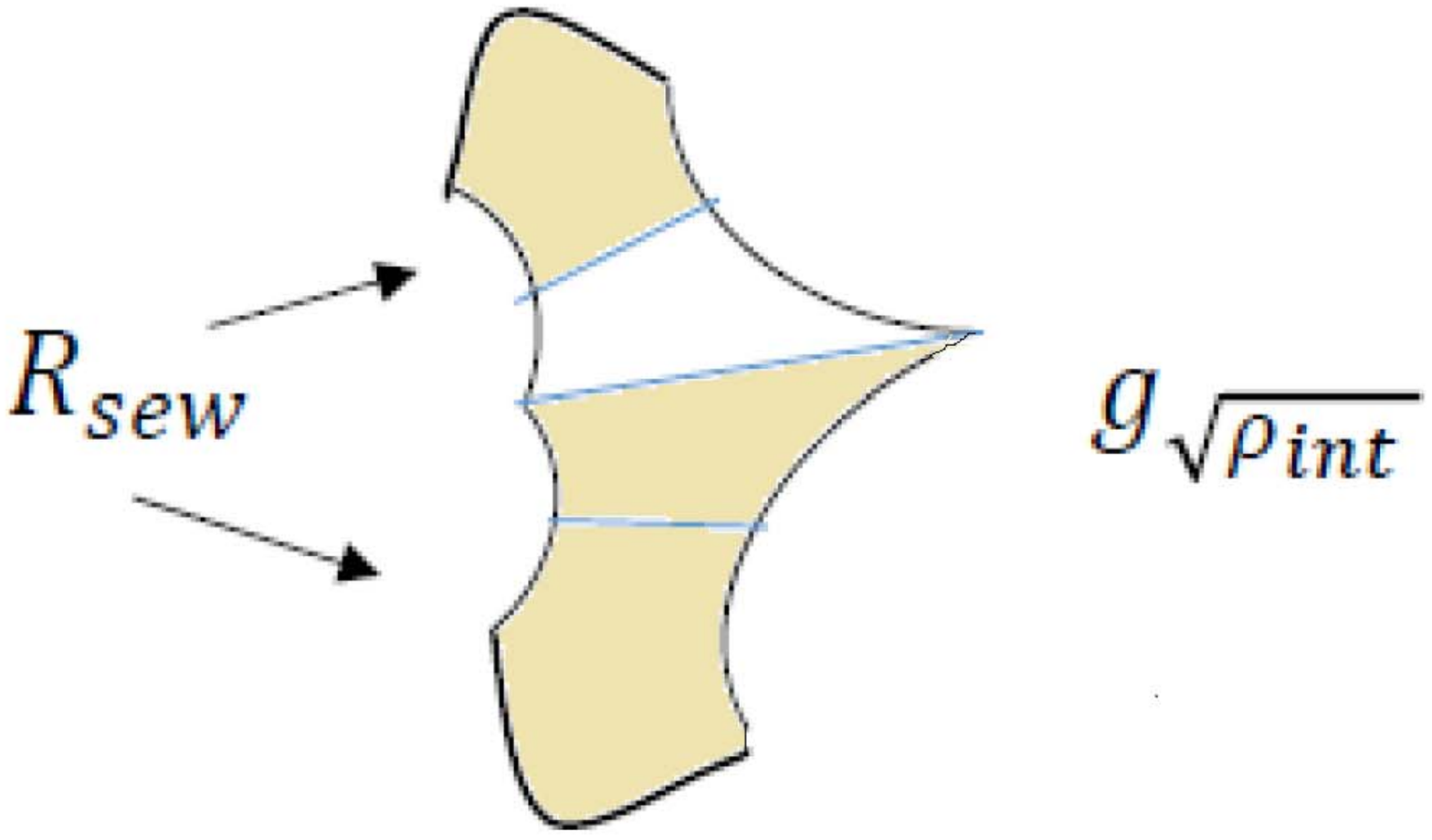}\\
\includegraphics[width =0.7\textwidth]{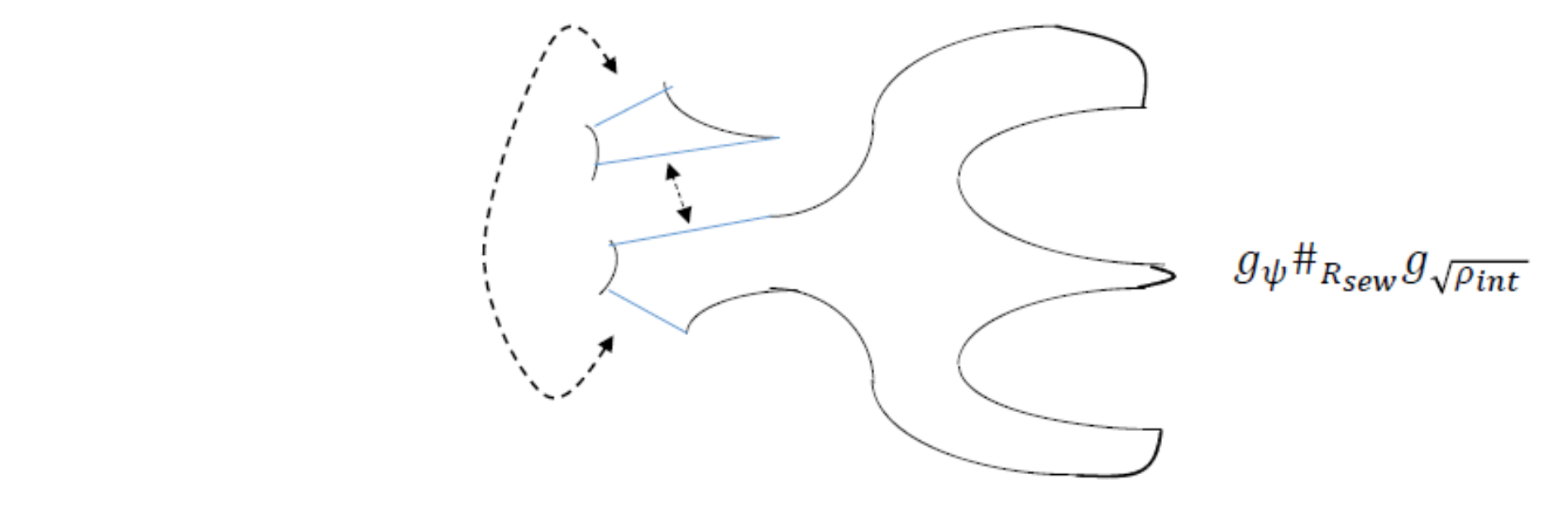} \ \ \\
\caption{A cut-and-paste on a single bulk geometry.  {\bf Top Row:} At left, two boundary regions $R_1,R_2$ and their RT surfaces $\gamma_{R_1}, \gamma_{R_2}$ are marked on a time-symmetric slice of a bulk geometry $g_\psi$. The wedges of $R_1,R_2$ are assumed to be CPT-conjugate.  At right, a new intermediate boundary region $R_{int}$ satisfying ${\bar R}_2 \supset R_{int} \supset R_1$ has been marked  along with its RT surface $\gamma_{R_{int}}$. {\bf Center Row:} On the boundary of $g_\psi$ we have defined $R_{sew} = R_{int} \cup R_2$ (left).  At right is the geometry $g_{\sqrt{\rho_{int}}}$ dual to the canonical purification $\sqrt{\rho_{R_{int}}}$ of $\rho_{R_{int}}$.  The upper half of its boundary is $R_{int}$, while the lower half defines ${\bar R}_{int}$.   Here we take $R_{sew} = {\bar R}_{int} \cup R_1$. Assuming conditions (ii) and (iii) from the main text, the entanglement wedge of $R_{sew}$ in $g_\psi$ (left) is the union of wedges for $R_{int}$ and $R_2$ (shaded regions), while in $g_{\sqrt{\rho_{int}}}$ (right) the entanglement wedge of $R_{sew}$ is the union of wedges for ${\bar R}_{int}$ and $R_1$ (also shaded).  The shaded regions in the two geometries are CPT-conjugate. {\bf Bottom:} Excising the entanglement wedge of $R_{sew}$ from both $g_\psi$ and $g_{\sqrt{\rho_{int}}}$ and gluing the remaining (unshaded) parts along $\gamma_{R_{sew}} = \gamma_{R_{int}} \cup \gamma_{R_{1,2}}$ gives $g_\psi \#_{R_{sew}} g_{\sqrt{\rho_{R_{int}}}}$.  The gluing along $\gamma_{R_{int}}$ simply reassembles the full wedge dual to ${\bar R}_1 \cap {\bar R}_2$ in $g_\psi$. Gluing along the remaining piece of $\gamma_{R_{sew}}$ identifies $\gamma_{R_1}$ with $\gamma_{R_2}$. }
\label{fig:SelfGlue}
\end{figure}

\noindent
to $\gamma_{R_2}$ as shown in the bottom line of figure \ref{fig:SelfGlue}.  We have thus shown the desired bulk geometry to be dual to $|\psi \#_{R_{sew}} \sqrt{\rho_{R_{int}}}\rangle$.

In the above discussion of gluing a state to itself, the vanishing leading-order mutual information between $R_{int}$ and $R_2$, and between ${\bar R}_{int}$ and $R_2$, in fact played two key roles. First, these conditions ensured that the states $\sqrt{\rho_{R_{int}}}$ and $|\psi\rangle$ were appropriately CPT-conjguate so that our results could be applied.  But even if this were somehow independently guaranteed, our algorithm amputates the bulk duals to both of these states along $\gamma_{R_{sew}}$ and then glues the remaining pieces together to define the final bulk spacetime.  When either of the above leading-order mutual informations are is non-trivial, the amputation step would remove a larger piece of the bulk, and the final geometry would be distinctly smaller than that obtained by simply gluing $\gamma_{R_1}$ to $\gamma_{R_2}$.

We will largely leave applications of our construction for future investigation.  However, before closing we briefly note that it can be used to implement the bulk gluing operations described in the second construction of \cite{Bao:2019zqc}, this giving a CFT state in which the so-called multipartite entanglement-wedge cross-section \cite{Bao:2018gck} can be realized as the entropy of a region of the CFT\footnote{The construction described in the original version of \cite{Bao:2019zqc} requires a generalization in the case where their $n$-party entanglement wedge has an HRT-surface with $k < n$ connected components.  We thank Ning Bao for correspondence on this issue.}.  In contrast, the first construction of \cite{Bao:2019zqc} involves gluing a state to itself.  As discussed above, in that case our procedure can be applied only when conditions (i), (ii), and (iii) are satisfied.  While we expect that this is in fact the case in the first construction of \cite{Bao:2019zqc}, we have not attempted to give an exhaustive proof.  Other applications remain to be investigated, but it would be especially interesting to understand if arguments like those above could be used to derive the surface-state correspondence of \cite{Miyaji:2015yva} or related results.

\paragraph{Acknowledgments}
It is a pleasure to thank Xi Dong, Daniel Harlow, Veronika Hubeny, Henry Maxfield, Mukund Rangamani, and Simon Ross for many years of discussions about cutting and pasting path integrals, geometries, and tensor networks.
This material is based upon work supported by the Air Force Office of Scientific Research under award number FA9550-19-1-0360.  The work was also supported in part by funds from the University of California.

\bibliographystyle{JHEP}

\bibliography{biblio}

\providecommand{\href}[2]{#2}\begingroup\raggedright\begin{thebibliography}{10}

\bibitem{Swingle:2009bg}
B.~Swingle, \emph{{Entanglement Renormalization and Holography}},
  \href{http://dx.doi.org/10.1103/PhysRevD.86.065007}{\emph{Phys. Rev.}
  {\bfseries D86} (2012) 065007},
  [\href{https://arxiv.org/abs/0905.1317}{{\ttfamily 0905.1317}}].

\bibitem{Qi:2013caa}
X.-L. Qi, \emph{{Exact holographic mapping and emergent space-time geometry}},
  \href{https://arxiv.org/abs/1309.6282}{{\ttfamily 1309.6282}}.

\bibitem{Evenbly}
G.~Evenbly and G.~Vidal, \emph{{Tensor network states and geometry}},
  \href{http://dx.doi.org/10.1007/s10955-011-0237-4}{\emph{J. Stat. Phys.}
  {\bfseries 145} (2012) 891},
  [\href{https://arxiv.org/abs/1106.1082}{{\ttfamily 1106.1082}}].

\bibitem{MolinaVilaplana:2011xt}
J.~Molina-Vilaplana and P.~Sodano, \emph{{Holographic View on Quantum
  Correlations and Mutual Information between Disjoint Blocks of a Quantum
  Critical System}},
  \href{http://dx.doi.org/10.1007/JHEP10(2011)011}{\emph{JHEP} {\bfseries 10}
  (2011) 011}, [\href{https://arxiv.org/abs/1108.1277}{{\ttfamily 1108.1277}}].

\bibitem{Swingle:2012wq}
B.~Swingle, \emph{{Constructing holographic spacetimes using entanglement
  renormalization}},  \href{https://arxiv.org/abs/1209.3304}{{\ttfamily
  1209.3304}}.

\bibitem{Matsueda:2012xm}
H.~Matsueda, M.~Ishihara and Y.~Hashizume, \emph{{Tensor network and a black
  hole}}, \href{http://dx.doi.org/10.1103/PhysRevD.87.066002}{\emph{Phys. Rev.}
  {\bfseries D87} (2013) 066002},
  [\href{https://arxiv.org/abs/1208.0206}{{\ttfamily 1208.0206}}].

\bibitem{Pastawski:2015qua}
F.~Pastawski, B.~Yoshida, D.~Harlow and J.~Preskill, \emph{{Holographic quantum
  error-correcting codes: Toy models for the bulk/boundary correspondence}},
  \href{http://dx.doi.org/10.1007/JHEP06(2015)149}{\emph{JHEP} {\bfseries 06}
  (2015) 149}, [\href{https://arxiv.org/abs/1503.06237}{{\ttfamily
  1503.06237}}].

\bibitem{Hayden:2016cfa}
P.~Hayden, S.~Nezami, X.-L. Qi, N.~Thomas, M.~Walter and Z.~Yang,
  \emph{{Holographic duality from random tensor networks}},
  \href{http://dx.doi.org/10.1007/JHEP11(2016)009}{\emph{JHEP} {\bfseries 11}
  (2016) 009}, [\href{https://arxiv.org/abs/1601.01694}{{\ttfamily
  1601.01694}}].

\bibitem{Czech:2012bh}
B.~Czech, J.~L. Karczmarek, F.~Nogueira and M.~Van~Raamsdonk, \emph{{The
  Gravity Dual of a Density Matrix}},
  \href{http://dx.doi.org/10.1088/0264-9381/29/15/155009}{\emph{Class. Quant.
  Grav.} {\bfseries 29} (2012) 155009},
  [\href{https://arxiv.org/abs/1204.1330}{{\ttfamily 1204.1330}}].

\bibitem{Bousso:2012mh}
R.~Bousso, B.~Freivogel, S.~Leichenauer, V.~Rosenhaus and C.~Zukowski,
  \emph{{Null Geodesics, Local CFT Operators and AdS/CFT for Subregions}},
  \href{http://dx.doi.org/10.1103/PhysRevD.88.064057}{\emph{Phys. Rev.}
  {\bfseries D88} (2013) 064057},
  [\href{https://arxiv.org/abs/1209.4641}{{\ttfamily 1209.4641}}].

\bibitem{Hubeny:2012wa}
V.~E. Hubeny and M.~Rangamani, \emph{{Causal Holographic Information}},
  \href{http://dx.doi.org/10.1007/JHEP06(2012)114}{\emph{JHEP} {\bfseries 06}
  (2012) 114}, [\href{https://arxiv.org/abs/1204.1698}{{\ttfamily 1204.1698}}].

\bibitem{Ryu:2006ef}
S.~Ryu and T.~Takayanagi, \emph{{Aspects of Holographic Entanglement Entropy}},
  \href{http://dx.doi.org/10.1088/1126-6708/2006/08/045}{\emph{JHEP} {\bfseries
  08} (2006) 045}, [\href{https://arxiv.org/abs/hep-th/0605073}{{\ttfamily
  hep-th/0605073}}].

\bibitem{Ryu:2006bv}
S.~Ryu and T.~Takayanagi, \emph{{Holographic derivation of entanglement entropy
  from AdS/CFT}},
  \href{http://dx.doi.org/10.1103/PhysRevLett.96.181602}{\emph{Phys. Rev.
  Lett.} {\bfseries 96} (2006) 181602},
  [\href{https://arxiv.org/abs/hep-th/0603001}{{\ttfamily hep-th/0603001}}].

\bibitem{Hubeny:2007xt}
V.~E. Hubeny, M.~Rangamani and T.~Takayanagi, \emph{{A Covariant holographic
  entanglement entropy proposal}},
  \href{http://dx.doi.org/10.1088/1126-6708/2007/07/062}{\emph{JHEP} {\bfseries
  07} (2007) 062}, [\href{https://arxiv.org/abs/0705.0016}{{\ttfamily
  0705.0016}}].

\bibitem{Miyaji:2015yva}
M.~Miyaji and T.~Takayanagi, \emph{{Surface/State Correspondence as a
  Generalized Holography}},
  \href{http://dx.doi.org/10.1093/ptep/ptv089}{\emph{PTEP} {\bfseries 2015}
  (2015) 073B03}, [\href{https://arxiv.org/abs/1503.03542}{{\ttfamily
  1503.03542}}].

\bibitem{Takayanagi:2017knl}
T.~Takayanagi and K.~Umemoto, \emph{{Entanglement of purification through
  holographic duality}},
  \href{http://dx.doi.org/10.1038/s41567-018-0075-2}{\emph{Nature Phys.}
  {\bfseries 14} (2018) 573--577},
  [\href{https://arxiv.org/abs/1708.09393}{{\ttfamily 1708.09393}}].

\bibitem{Nguyen:2017yqw}
P.~Nguyen, T.~Devakul, M.~G. Halbasch, M.~P. Zaletel and B.~Swingle,
  \emph{{Entanglement of purification: from spin chains to holography}},
  \href{http://dx.doi.org/10.1007/JHEP01(2018)098}{\emph{JHEP} {\bfseries 01}
  (2018) 098}, [\href{https://arxiv.org/abs/1709.07424}{{\ttfamily
  1709.07424}}].

\bibitem{Tamaoka:2018ned}
K.~Tamaoka, \emph{{Entanglement Wedge Cross Section from the Dual Density
  Matrix}}, \href{http://dx.doi.org/10.1103/PhysRevLett.122.141601}{\emph{Phys.
  Rev. Lett.} {\bfseries 122} (2019) 141601},
  [\href{https://arxiv.org/abs/1809.09109}{{\ttfamily 1809.09109}}].

\bibitem{Kudler-Flam:2018qjo}
J.~Kudler-Flam and S.~Ryu, \emph{{Entanglement negativity and minimal
  entanglement wedge cross sections in holographic theories}},
  \href{http://dx.doi.org/10.1103/PhysRevD.99.106014}{\emph{Phys. Rev.}
  {\bfseries D99} (2019) 106014},
  [\href{https://arxiv.org/abs/1808.00446}{{\ttfamily 1808.00446}}].

\bibitem{Dutta:2019gen}
S.~Dutta and T.~Faulkner, \emph{{A canonical purification for the entanglement
  wedge cross-section}},  \href{https://arxiv.org/abs/1905.00577}{{\ttfamily
  1905.00577}}.

\bibitem{Umemoto:2018jpc}
K.~Umemoto and Y.~Zhou, \emph{{Entanglement of Purification for Multipartite
  States and its Holographic Dual}},
  \href{http://dx.doi.org/10.1007/JHEP10(2018)152}{\emph{JHEP} {\bfseries 10}
  (2018) 152}, [\href{https://arxiv.org/abs/1805.02625}{{\ttfamily
  1805.02625}}].

\bibitem{Bao:2017nhh}
N.~Bao and I.~F. Halpern, \emph{{Holographic Inequalities and Entanglement of
  Purification}}, \href{http://dx.doi.org/10.1007/JHEP03(2018)006}{\emph{JHEP}
  {\bfseries 03} (2018) 006},
  [\href{https://arxiv.org/abs/1710.07643}{{\ttfamily 1710.07643}}].

\bibitem{Hirai:2018jwy}
H.~Hirai, K.~Tamaoka and T.~Yokoya, \emph{{Towards Entanglement of Purification
  for Conformal Field Theories}},
  \href{http://dx.doi.org/10.1093/ptep/pty063}{\emph{PTEP} {\bfseries 2018}
  (2018) 063B03}, [\href{https://arxiv.org/abs/1803.10539}{{\ttfamily
  1803.10539}}].

\bibitem{Bao:2018gck}
N.~Bao and I.~F. Halpern, \emph{{Conditional and Multipartite Entanglements of
  Purification and Holography}},
  \href{http://dx.doi.org/10.1103/PhysRevD.99.046010}{\emph{Phys. Rev.}
  {\bfseries D99} (2019) 046010},
  [\href{https://arxiv.org/abs/1805.00476}{{\ttfamily 1805.00476}}].

\bibitem{Bao:2018fso}
N.~Bao, A.~Chatwin-Davies and G.~N. Remmen, \emph{{Entanglement of Purification
  and Multiboundary Wormhole Geometries}},
  \href{http://dx.doi.org/10.1007/JHEP02(2019)110}{\emph{JHEP} {\bfseries 02}
  (2019) 110}, [\href{https://arxiv.org/abs/1811.01983}{{\ttfamily
  1811.01983}}].

\bibitem{Agon:2018lwq}
C.~A. Agón, J.~De~Boer and J.~F. Pedraza, \emph{{Geometric Aspects of
  Holographic Bit Threads}},
  \href{http://dx.doi.org/10.1007/JHEP05(2019)075}{\emph{JHEP} {\bfseries 05}
  (2019) 075}, [\href{https://arxiv.org/abs/1811.08879}{{\ttfamily
  1811.08879}}].

\bibitem{Caputa:2018xuf}
P.~Caputa, M.~Miyaji, T.~Takayanagi and K.~Umemoto, \emph{{Holographic
  Entanglement of Purification from Conformal Field Theories}},
  \href{http://dx.doi.org/10.1103/PhysRevLett.122.111601}{\emph{Phys. Rev.
  Lett.} {\bfseries 122} (2019) 111601},
  [\href{https://arxiv.org/abs/1812.05268}{{\ttfamily 1812.05268}}].

\bibitem{Kudler-Flam:2019oru}
J.~Kudler-Flam, I.~MacCormack and S.~Ryu, \emph{{Holographic entanglement
  contour, bit threads, and the entanglement tsunami}},
  \href{http://dx.doi.org/10.1088/1751-8121/ab2dae}{\emph{J. Phys.} {\bfseries
  A52} (2019) 325401}, [\href{https://arxiv.org/abs/1902.04654}{{\ttfamily
  1902.04654}}].

\bibitem{Du:2019emy}
D.-H. Du, C.-B. Chen and F.-W. Shu, \emph{{Bit threads and holographic
  entanglement of purification}},
  \href{http://dx.doi.org/10.1007/JHEP08(2019)140}{\emph{JHEP} {\bfseries 08}
  (2019) 140}, [\href{https://arxiv.org/abs/1904.06871}{{\ttfamily
  1904.06871}}].

\bibitem{Jokela:2019ebz}
N.~Jokela and A.~Pönni, \emph{{Notes on entanglement wedge cross sections}},
  \href{http://dx.doi.org/10.1007/JHEP07(2019)087}{\emph{JHEP} {\bfseries 07}
  (2019) 087}, [\href{https://arxiv.org/abs/1904.09582}{{\ttfamily
  1904.09582}}].

\bibitem{Bao:2019wcf}
N.~Bao, A.~Chatwin-Davies, J.~Pollack and G.~N. Remmen, \emph{{Towards a Bit
  Threads Derivation of Holographic Entanglement of Purification}},
  \href{http://dx.doi.org/10.1007/JHEP07(2019)152}{\emph{JHEP} {\bfseries 07}
  (2019) 152}, [\href{https://arxiv.org/abs/1905.04317}{{\ttfamily
  1905.04317}}].

\bibitem{Harper:2019lff}
J.~Harper and M.~Headrick, \emph{{Bit threads and holographic entanglement of
  purification}}, \href{http://dx.doi.org/10.1007/JHEP08(2019)101}{\emph{JHEP}
  {\bfseries 08} (2019) 101},
  [\href{https://arxiv.org/abs/1906.05970}{{\ttfamily 1906.05970}}].

\bibitem{Kusuki:2019zsp}
Y.~Kusuki, J.~Kudler-Flam and S.~Ryu, \emph{{Derivation of holographic
  negativity in ${\it AdS}_3/{\it CFT}_2$}},
  \href{https://arxiv.org/abs/1907.07824}{{\ttfamily 1907.07824}}.

\bibitem{Levin:2019krg}
J.~Levin, O.~DeWolfe and G.~Smith, \emph{{Correlation measures and distillable
  entanglement in AdS/CFT}},
  \href{https://arxiv.org/abs/1909.04727}{{\ttfamily 1909.04727}}.

\bibitem{Kusuki:2019evw}
Y.~Kusuki and K.~Tamaoka, \emph{{Entanglement Wedge Cross Section from CFT:
  Dynamics of Local Operator Quench}},
  \href{https://arxiv.org/abs/1909.06790}{{\ttfamily 1909.06790}}.

\bibitem{Akers:2018fow}
C.~Akers and P.~Rath, \emph{{Holographic Renyi Entropy from Quantum Error
  Correction}}, \href{http://dx.doi.org/10.1007/JHEP05(2019)052}{\emph{JHEP}
  {\bfseries 05} (2019) 052},
  [\href{https://arxiv.org/abs/1811.05171}{{\ttfamily 1811.05171}}].

\bibitem{Dong:2018seb}
X.~Dong, D.~Harlow and D.~Marolf, \emph{{Flat entanglement spectra in
  fixed-area states of quantum gravity}},
  \href{https://arxiv.org/abs/1811.05382}{{\ttfamily 1811.05382}}.

\bibitem{Donnelly:2016qqt}
W.~Donnelly, B.~Michel, D.~Marolf and J.~Wien, \emph{{Living on the Edge: A Toy
  Model for Holographic Reconstruction of Algebras with Centers}},
  \href{http://dx.doi.org/10.1007/JHEP04(2017)093}{\emph{JHEP} {\bfseries 04}
  (2017) 093}, [\href{https://arxiv.org/abs/1611.05841}{{\ttfamily
  1611.05841}}].

\bibitem{Lewkowycz:2013nqa}
A.~Lewkowycz and J.~Maldacena, \emph{{Generalized gravitational entropy}},
  \href{http://dx.doi.org/10.1007/JHEP08(2013)090}{\emph{JHEP} {\bfseries 08}
  (2013) 090}, [\href{https://arxiv.org/abs/1304.4926}{{\ttfamily 1304.4926}}].

\bibitem{Dong:2016hjy}
X.~Dong, A.~Lewkowycz and M.~Rangamani, \emph{{Deriving covariant holographic
  entanglement}}, \href{http://dx.doi.org/10.1007/JHEP11(2016)028}{\emph{JHEP}
  {\bfseries 11} (2016) 028},
  [\href{https://arxiv.org/abs/1607.07506}{{\ttfamily 1607.07506}}].

\bibitem{Harlow:2016vwg}
D.~Harlow, \emph{{The Ryu–Takayanagi Formula from Quantum Error Correction}},
  \href{http://dx.doi.org/10.1007/s00220-017-2904-z}{\emph{Commun. Math. Phys.}
  {\bfseries 354} (2017) 865--912},
  [\href{https://arxiv.org/abs/1607.03901}{{\ttfamily 1607.03901}}].

\bibitem{DMtoappear}
X.~Dong and D.~Marolf, ``One-loop universality of holographic codes.'' 2019.

\bibitem{Engelhardt:2018kcs}
N.~Engelhardt and A.~C. Wall, \emph{{Coarse Graining Holographic Black Holes}},
  \href{http://dx.doi.org/10.1007/JHEP05(2019)160}{\emph{JHEP} {\bfseries 05}
  (2019) 160}, [\href{https://arxiv.org/abs/1806.01281}{{\ttfamily
  1806.01281}}].

\bibitem{Faulkner:2017vdd}
T.~Faulkner and A.~Lewkowycz, \emph{{Bulk locality from modular flow}},
  \href{http://dx.doi.org/10.1007/JHEP07(2017)151}{\emph{JHEP} {\bfseries 07}
  (2017) 151}, [\href{https://arxiv.org/abs/1704.05464}{{\ttfamily
  1704.05464}}].

\bibitem{Balasubramanian:2014hda}
V.~Balasubramanian, P.~Hayden, A.~Maloney, D.~Marolf and S.~F. Ross,
  \emph{{Multiboundary Wormholes and Holographic Entanglement}},
  \href{http://dx.doi.org/10.1088/0264-9381/31/18/185015}{\emph{Class. Quant.
  Grav.} {\bfseries 31} (2014) 185015},
  [\href{https://arxiv.org/abs/1406.2663}{{\ttfamily 1406.2663}}].

\bibitem{Marolf:2015vma}
D.~Marolf, H.~Maxfield, A.~Peach and S.~F. Ross, \emph{{Hot multiboundary
  wormholes from bipartite entanglement}},
  \href{http://dx.doi.org/10.1088/0264-9381/32/21/215006}{\emph{Class. Quant.
  Grav.} {\bfseries 32} (2015) 215006},
  [\href{https://arxiv.org/abs/1506.04128}{{\ttfamily 1506.04128}}].

\bibitem{Bao:2019zqc}
N.~Bao and N.~Cheng, \emph{{Multipartite Reflected Entropy}},
  \href{https://arxiv.org/abs/1909.03154}{{\ttfamily 1909.03154}}.

\end{thebibliography}\endgroup

\end{document}